\pdfoutput=1
\documentclass[usenatbib,onecolumn]{mnras}

\usepackage[T1]{fontenc}
\usepackage{ae,aecompl}

\usepackage[dvipdfmx]{graphicx}	
\usepackage{amsmath}	
\usepackage{amssymb}	
\usepackage{color}
\usepackage{dcolumn}
\usepackage{bm}
\usepackage{comment}

\newcommand{\beq}{\begin{equation}}
\newcommand{\beqa}{\begin{eqnarray}}
\newcommand{\eeq}{\end{equation}}
\newcommand{\eeqa}{\end{eqnarray}}

\newcommand{\simgt}{\lower.5ex\hbox{$\; \buildrel > \over \sim \;$}}
\newcommand{\simlt}{\lower.5ex\hbox{$\; \buildrel < \over \sim \;$}}

\newcommand{\bd}[1]{\mbox{\boldmath $#1$}}

\newcommand{\MS}[1]{\textcolor{black}{#1}}

\RequirePackage{lineno}

\def\aj{{AJ}}                   
\def\apj{{ApJ}}                 
\def\apjl{{ApJ}}                
\def\apjs{{ApJS}}               
\def\ao{{Appl.~Opt.}}           
\def\aap{{A\&A}}                

\def\jcap{{J. Cosmology Astropart. Phys.}}
\def\mnras{MNRAS}             
\def\prd{{Phys.~Rev.~D}}        
\def\pasj{{PASJ}}               
\title[Weak lensing halo-shear-shear correlation]{
Probing the shape and internal structure of dark matter halos with 
the halo-shear-shear three-point correlation function
}

\author[M.~Shirasaki and N.~Yoshida]
{Masato Shirasaki$^{1}$
\thanks{E-mail:masato.shirasaki@nao.ac.jp}
and Naoki Yoshida$^{2,3,4,5}$
\\
$^{1}$Division of Theoretical Astronomy, National Astronomical Observatory of Japan, 2-21-1 Osawa, Mitaka, Tokyo 181-8588, Japan\\
$^{2}$Department of Physics, University of Tokyo, 7-3-1 Hongo, Bunkyo,
Tokyo 113-0033, Japan\\
$^{3}$Research Center for the Early Universe, University of Tokyo, 7-3-1 Hongo,
Bunkyo, Tokyo 113-0033, Japan \\
$^{4}$Kavli Institute for the Physics and Mathematics of the 
Universe (WPI), UTIAS, \\
University of Tokyo, 5-1-5 Kashiwanoha, Kashiwa 277-85
83, Japan\\
$^{5}$CREST, JST, 4-1-8 Honcho, Kawaguchi, Saitama, 332-0012, 
Japan \\
}

\pubyear{2017}

\begin{document}
\label{firstpage}
\pagerange{\pageref{firstpage}--\pageref{lastpage}}
\maketitle

\begin{abstract}
  Weak lensing three-point statistics are 
  powerful probes of the structure of dark matter halos.
  We propose to use the correlation of the positions of galaxies
  with the shapes of background galaxy pairs,
  known as the halo-shear-shear correlation (HSSC),
  to measure the mean halo ellipticity
  and the abundance of subhalos in a statistical manner.
  We run high-resolution cosmological $N$-body simulations and use the
  outputs to measure the HSSC for galaxy halos and cluster halos.
  Non-spherical halos cause a characteristic azimuthal
  variation of the HSSC, and massive subhalos in the outer region
  near the virial radius contribute to $\sim10\%$ of the HSSC amplitude. 
  \MS{Using the HSSC and its covariance estimated from our $N$-body simulations,} 
  we make forecast for constraining the internal structure of
  dark matter halos with future galaxy surveys.
  With 1000 galaxy groups with mass greater than $10^{13.5}\, h^{-1}M_{\odot}$, 
  the average halo ellipticity can be measured with
  an accuracy of ten percent. A spherical, smooth mass distribution
  can be ruled out at a $\sim5\sigma$ significance level.
  The existence of subhalos whose masses are in 1-10 percent of the
  main halo mass can be detected
  with $\sim10^4$ galaxies/clusters. We conclude that the HSSC
  provides valuable information on the structure of dark
  halos and hence on the nature of dark matter.
\end{abstract}


\section{INTRODUCTION}
An array of astronomical observations suggest
that ordinary matter accounts for only about twenty percent
of the mass content of the Universe.
The rest consists of some unknown substitute called dark matter.
Measurements of temperature anisotropies
in the cosmic microwave background (CMB)
suggest the existence of dark matter at 
$\sim50\sigma$ significance level,
and also that dark matter should be non-baryonic
\citep[e.g.][]{2016A&A...594A..13P}.
Although the nature of dark matter is still unknown,
currently available astronomical data of
large-scale structure of
the comoving lengthscale of 
$\simgt10\, {\rm Mpc}$ are consistent
with a simple model of cold dark matter (CDM)
that posits that dark matter is made of 
stable, collisionless, and non-relativistic particles
\citep[e.g.][]{2006PhRvD..74l3507T}.

There appear to exist a few discrepancies between
the theoretical prediction of the standard CDM model
and observations of galactic-size objects 
\citep[see, e.g.][for a short review]{2015PNAS..11212249W}.
Such a ``small-scale crisis'' of the CDM model
can be resolved by modifying the particle nature of dark matter.
For example, in warm dark matter (WDM) models,
the primordial density perturbations at small length scales
are suppressed owing to free-streaming of dark matter particles
\citep[e.g.][]{1994PhRvL..72...17D}.
Self-interacting dark matter (SIDM) model is another 
alternative. If dark matter is made of collisional particles \citep[e.g.][]{2000PhRvL..84.3760S},
dark halos are rounder, and typically have a density profile with a constant density ``core" that appears to be in agreement with observations of rotation curves
of galaxies \citep[e.g.][]{2016PhRvL.116d1302K}.
A promising way to discriminate the variants of dark matter models
is to directly probe the structure of 
dark matter halos.

Gravitational lensing provides a unique physical method 
to probe matter distribution in and around galaxies.
Although the lensing effect on individual source galaxies
is expected to be small, 
statistical analyses of a large number of sources 
enable us to reconstruct the matter density distribution 
in an unbiased way.
In particular, cross-correlation of large-scale structure tracers such as galaxies and galaxy clusters with shapes of background galaxies, 
measured by a stacking method, can be used to 
infer the average total matter distribution 
\MS{
around galaxies \citep[e.g.][]{
1996ApJ...466..623B, 1998ApJ...503..531H, 2002MNRAS.335..311G,
2004ApJ...606...67H, 2006MNRAS.368..715M}
and galaxy clusters \citep[e.g.][]{
2006MNRAS.372..758M, 2007arXiv0709.1159J,
2013ApJ...769L..35O, 2014ApJ...784L..25C}.
The gravitational lensing effects have been detected and extensively studied also for individual galaxy clusters \citep[e.g.][]{2005ApJ...619L.143B, 2012MNRAS.420.3213O, 
2012MNRAS.427.1298H, 2014MNRAS.439...48A, 2016JCAP...08..013B}.
}
Such measurements play a central role in constraining
the \MS{spherically averaged} density profile, but it is difficult to extract more information about the fine structure of dark matter halos.
Numerical simulations predict 
that dark matter halos are triaxial \citep[e.g.][]{2002ApJ...574..538J}
and have abundant substructures \citep[e.g.][]{1998MNRAS.299..728T}.
Recent simulations for a variant of dark matter models
show that the structure of dark halos 
are useful to constrain the nature of dark matter 
\citep[e.g.][]{2013MNRAS.430..105P, 2014MNRAS.444.2333E}.

\MS{
The shape of dark matter halos has been
investigated with stacked weak lensing analysis so far.
There exist various studies to measure the projected ellipticity of surface mass density around galaxy-sized halos \citep[e.g.][]{2004ApJ...606...67H, 2006MNRAS.370.1008M, 2007ApJ...669...21P, 2012A&A...545A..71V, 2015MNRAS.454.1432S},
whereas recent galaxy-imaging observations allow us
to study the shape of more massive halos, galaxy groups \citep[][]{2016MNRAS.457.4135C, 2017MNRAS.467.4131V}
and galaxy clusters \citep[][]{2009ApJ...695.1446E}.
Most previous lensing measurements of the halo shape require some proxy of orientation of the principal axes in surface mass density 
around lensing objects\footnote{\MS{\citet{2010MNRAS.405.2215O} have studied the two-dimensional (2D) weak lensing signals around galaxy clusters on an individual object basis. Their approach does not need a proxy of halo orientation to measure the halo shape, but it requires very deep imaging data to increase the signal-to-noise ratio. In addition, the 2D fitting method seems to be applicable only to most massive galaxy clusters.}}.
On galaxy scales, it is commonly assumed that a galaxy resides
perfectly at the center of the dark halo.
For more massive objects, such as galaxy groups and clusters, one may expect that the distribution of member galaxies will follow the mass distribution.
However, there also exists some observational evidence
that central massive galaxies and their dark halos are not aligned \citep[][]{2009ApJ...694..214O}
and that there is only weak correlation between the dark mass ellipticity and that of the 
distribution of the member galaxies \citep[][]{2010MNRAS.405.2215O}.
Interestingly, recent hydrodynamics simulations show 
that the light distribution does not follow the underlying mass distribution in halos of galaxies and clusters \citep[e.g.][]{2015MNRAS.453..721V,2017MNRAS.472.1163C}.
It is still difficult to interpret the results of
conventional stacked weak lensing analysis appropriately,
without knowing the galaxy-halo misalignment a priori.
}

In this paper, we extend the method of stacked lensing to probe the internal structure of dark matter halos.
We consider a three-point correlation  
defined by the correlation of the shapes of background-galaxy pairs around the positions of lensing galaxies (halos).
Throughout this paper, we refer to this three-point correlation as halo-shear-shear correlation (HSSC).
\MS{
A great advantage of using the HSSC is that statistical measurement
can be done without relying on the light distribution of central galaxies or on the spatial distribution of the member galaxies.
}
\citet{2012A&A...548A.102S} use an isolated lens model to show that the HSSC, without using luminous tracers, can probe the halo ellipticity. More recently, \citet{2015JCAP...01..009A}
propose a statistical method to measure the HSSC and to constrain the mean ellipticity of galaxy-size halos
\MS{without using any information about the halo orientation}.
\MS{
The HSSC can be applied to study more massive objects 
like galaxy groups and clusters, and it will be more successful than galaxy-size halos since the lensing signal is expected to be stronger.
Nevertheless, it still remains uncertain if the HSSC can probe the average ellipticity of group and cluster halos in practice. The standard CDM model predicts that numerous dark substructures 
are expected to exist. They can induce possible anisotropic signals in the HSSC.
In addition, projected uncorrelated structures along a line
of sight can also affect the observed HSSC.
To study these effects on the HSSC simultaneously, 
we use high-resolution $N$-body simulations as one of the best approaches at present.
}
Observationally, HSSC has been already detected with a high significance by recent weak lensing surveys \citep[e.g.][]{2008A&A...479..655S, 2013MNRAS.430.2476S}, 
\MS{
but these measurements have still focused on 
the large-scale three-point correlation. 
Since ongoing and upcoming galaxy surveys 
hold promise
for measuring the HSSC with a high statistical significance
on scales down to virial radii of group- and cluster-sized halos,
it is important and timely to develop
an accurate theoretical model of the HSSC induced by
the internal structures of dark halos.
}
For this purpose, we use realistic halo catalogs from high-resolution N-body simulations to examine the ability of HSSC to detect or measure halo ellipticity.
We also study the impact of substructures
\MS{and projected large-scale structures along a line of sight}
on the HSSC for the first time.
Note that our analysis focuses on 
the HSSC for massive galaxies and galaxy clusters
and hence is complementary to the work of 
\citet{2015JCAP...01..009A}.

This paper is organized as follows. 
In Section~\ref{sec:hss}, we summarize the basics of weak lensing and lensing statistics used in this paper.
We explain the details of our lensing simulation 
and the methodology to estimate lensing statistics 
in Section~\ref{sec:sim_and_ana}.
In Section~\ref{sec:res}, 
\MS{we show how the HSSC will be sensitive to halo ellipticity
and presence of substructures, 
and study the projection effect along a line of sight.}
We then quantify the information content in the HSSC
on the internal structure of dark matter halos.
Conclusions and discussions are presented in Section~\ref{sec:con}.

\section{HALO-SHEAR-SHEAR CORRELATION}
\label{sec:hss}

\subsection{Weak lensing}
We first summarize the basics of gravitational lensing 
induced by large-scale structure \MS{\citep[also see e.g.,][for a thorough review]{2001PhR...340..291B}}.
Weak gravitational lensing effect is characterized by
the distortion of image of a source object by the 
following $2\times2$ matrix:
\beqa
A_{ij} = \frac{\partial x_{\rm true}^{i}}{\partial x_{\rm obs}^{j}}
           \equiv \left(
\begin{array}{cc}
1-\kappa -\gamma_{1} & -\gamma_{2}-\omega  \\
-\gamma_{2}+\omega & 1-\kappa+\gamma_{1} \\
\end{array}
\right), \label{distortion_tensor}
\eeqa
where we denote the observed position of a source object as $\bd{x}_{\rm obs}$ 
and the true position as $\bd{x}_{\rm true}$.
In the above equation, $\kappa$ is the convergence, $\gamma$ is the shear,
and $\omega$ is the rotation.
For a given surface mass density along a line of sight, 
convergence $\kappa$ is computed as 
\beqa
\kappa({\bd \theta}) = \frac{\Sigma(\bd \theta)}{\Sigma_{\rm crit}},
\label{eq:Sigma2kappa}
\eeqa 
where 
$\Sigma(\bd \theta)$ represents the surface mass density,
$\bd \theta$ is the impact vector in the lens plane,
and $\Sigma_{\rm crit}$ is known as the critical density defined by the following relation
\beqa
\Sigma_{\rm crit} = \frac{c^2}{4\pi G}\frac{D_{\rm s}}{D_{\rm l}D_{\rm ls}},
\eeqa
where $D_{\rm s}$, $D_{\rm l}$, and $D_{\rm ls}$ are the angular diameter distance
to the source, to the lens, and between the source and the lens, respectively.
Throughout this paper, we consider a single source redshift 
$z_s = 1$ for lensing calculations.

In optical weak lensing surveys, galaxy shapes
are commonly used as an estimator of lensing shear $\gamma$.
One can relate $\gamma$ with $\kappa$ by introducing the deflection potential $\psi$:
\beqa
\kappa &=& \frac{1}{2}\nabla_{c}\nabla_{c}^{*} \psi, \label{eq:kap2psi}\\
\gamma &=& \frac{1}{2}\nabla_{c}\nabla_{c} \psi \label{eq:gam2psi},
\eeqa
where $\gamma=\gamma_{1}+{\rm i} \gamma_{2}$
and $\nabla_{c} = \partial/\partial \theta_{x} + {\rm i} \partial/\partial \theta_{y}$.
In polar coordinates, the differential operator is written as
\beqa
\nabla_{c} = {\rm e}^{{\rm i}\varphi}
\left(\frac{\partial}{\partial \theta}+\frac{\rm i}{\theta}\frac{\partial}{\partial \varphi}\right),
\label{eq:def_ope}
\eeqa
where $\theta=|\bd \theta|$ and $\varphi = \tan^{-1}(\theta_{y}/\theta_{x})$.
It is useful to decompose the tangential and cross components
of the shear with respect to the lens center as
\beqa
\gamma_{t} +{\rm i}\gamma_{\times} = -{\rm e}^{-2{\rm i}\varphi}\gamma.
\label{eq:she_t_x}
\eeqa
With Eqs.~(\ref{eq:kap2psi}) -- (\ref{eq:she_t_x}), 
one can find
\beqa
\kappa(\theta, \varphi) 
&=& \frac{1}{2}\left(
\frac{\partial^2\psi}{\partial \theta^2}+
\frac{1}{\theta}\frac{\partial \psi}{\partial \theta}+
\frac{1}{\theta^2}\frac{\partial^2\psi}{\partial \varphi^2}
\right), \label{eq:kap_polar} \\ 
\gamma_{t}(\theta, \varphi)
&=& \frac{1}{2}\left(
-\frac{\partial^2\psi}{\partial \theta^2}+
\frac{1}{\theta}\frac{\partial \psi}{\partial \theta}+
\frac{1}{\theta^2}\frac{\partial^2\psi}{\partial \varphi^2}
\right), \label{eq:she_t_polar} \\
\gamma_{\times}(\theta, \varphi) 
&=& 
\frac{1}{\theta^2} \frac{\partial \psi}{\partial \varphi}-
\frac{1}{\theta} \frac{\partial^2 \psi}{\partial \theta \partial \varphi}.
\label{eq:she_x_polar}
\eeqa

\subsection{Estimator}

The azimuthal average of Eqs~(\ref{eq:kap_polar}) and (\ref{eq:she_t_polar}) yields \MS{\citep[e.g.,][]{1991ApJ...370....1M, 1993ApJ...404..441K}}
\beqa
\gamma_{t, 0} (\theta)
= \frac{2}{\theta^2}\int_{0}^{\theta} \theta^{\prime}\, 
\kappa_{0} (\theta^{\prime}) \, {\rm d}\theta^{\prime}
- \kappa_{0} (\theta). \label{eq:she_t_0}
\eeqa
Here, we use subscript 0 such that $F_{0}(\theta)$ represents the averaging operation over azimuthal angle $\varphi$ 
for a given two-dimensional field $F(\theta, \varphi)$.
A common estimator of $\gamma_{t, 0}$ is 
the cross-correlation function between the position of lenses and the tangential ellipticity of sources \MS{\citep[e.g.,][]{2000AJ....120.1198F,2001ApJ...554..881S}},
which is defined as 
\beqa
\xi_{h+}(\theta) = \langle n_{\rm halo}({\bd \alpha}) 
\epsilon_{t}({\bd \alpha}+{\bd \theta}) \rangle, \label{eq:ggl}
\eeqa
where $n_{\rm halo}$ is the number density of lens halos
and $\epsilon_{t}$ is the tangential component of source ellipticity with respect to the position of lenses.
\MS{We have introduced the vector ${\bd \alpha}$
to denote a position in the lens plane, and ${\bd \theta}$
represents the separation angle between the target lens and source.}

In this paper, we generalize Eq.~(\ref{eq:ggl})
to study the statistical properties of surface mass density.
A simple extension of Eq.~(\ref{eq:ggl}) is the three-point correlation function
between $n_{\rm halo}$ and $\epsilon_{t}$ given by
\beqa
\zeta_{h++}(\theta_{1}, \theta_{2}, \phi)
= \langle n_{\rm halo}({\bd \alpha}) 
\epsilon_{t}({\bd \alpha}+{\bd \theta}_{1})
\epsilon_{t}({\bd \alpha}+{\bd \theta}_{2})
\rangle
\label{eq:gggl}
\eeqa
where
\MS{
${\bd \theta}_{1} ({\bd \theta}_{2})$
represents the separation angle between the target lens at ${\bd \alpha}$ and the position of the first (second) source 
(galaxy).}
Note that $\theta_{1} = |{\bd \theta}_{1}|$, 
$\theta_{2} = |{\bd \theta}_{2}|$,
and $\phi = \cos^{-1} \left({\bd \theta}_{1} \cdot {\bd \theta}_{2}/\theta_{1}/\theta_{2}\right)$.
The estimator of Eq.~(\ref{eq:gggl}) 
is useful to 
extract the information of 
{\it asphericity} and 
{\it azimuthal asymmetry}
of surface mass density at the lens plane
\MS{(see Appendix~\ref{subsec:toy})}.
\MS{For comparison, we also consider  
the three-point correlation function in terms of convergence field $\kappa$,
which is defined by
\beqa
\zeta_{h\kappa\kappa}(\theta_{1}, \theta_{2}, \phi)
= \langle n_{\rm halo}({\bd \alpha}) 
\kappa({\bd \alpha}+{\bd \theta}_{1})
\kappa({\bd \alpha}+{\bd \theta}_{2})
\rangle
\label{eq:gggl_kappa}.
\eeqa
Hereafter, we refer to $\zeta_{h\kappa\kappa}$ 
as halo-kappa-kappa correlation (HKKC).}

\section{SIMULATION AND ANALYSIS}
\label{sec:sim_and_ana}
\subsection{$N$-body simulation}
\label{subsec:sim}
We run cosmological $N$-body simulations to generate
a set of three-dimensional matter density field. 
We use the parallel Tree-Particle Mesh 
code {\tt Gadget2}
\citep{2005MNRAS.364.1105S}. 
We employ $1024^3$ dark matter 
particles in a volume of $200\, h^{-1}$Mpc on a side. 
We generate the initial conditions using a parallel code 
developed by \citet{2009PASJ...61..321N} and
\citet{2011A&A...527A..87V}, which employs the 
second-order Lagrangian perturbation theory 
\cite[e.g.,][]{2006MNRAS.373..369C}.
The initial redshift is set to 
$z_{\rm init}=49$, where we compute the linear matter transfer
function using {\tt CAMB} \citep{2000ApJ...538..473L}.
Our fiducial cosmology adopts the following parameters:
present-day matter density parameter  $\Omega_{\rm m0}=0.272$, dark energy density $\Omega_{\Lambda 0}=0.728$, 
the density fluctuation amplitude
$\sigma_{8}=0.809$,
the parameter of the equation of state of dark energy $w_{0} = -1$,
Hubble parameter $h=0.704$ and 
the scalar spectral index $n_s=0.963$.
These parameters are consistent with 
the \textit{WMAP} seven-year results \citep{2011ApJS..192...18K}.
\MS{In our simulation, the particle mass is set to be $5.6\times10^{8}\, h^{-1}M_{\odot}$,
and we set the softening length to be $5.85\, h^{-1}$kpc, corresponding
to about three percent of the mean separation length of the particles.}

In the output of the $N$-body simulations, 
we locate dark matter halos using the standard friend-of-friend (FOF) algorithm
with the linking parameter of $0.2$.
We define the mass of each halo 
\MS{by using the spherical overdensity mass with $\Delta=200$ 
with respect to the mean matter density.
We denote this mass $M_{\rm 200m}$.}
\MS{In the following analysis, 
we use halos with mass greater 
than $M_{\rm 200m}\ge10^{13.5}\, h^{-1}M_{\odot}$
at the redshift of 0.32, which is the typical redshift of lens objects
in optical weak-lensing surveys. We analyse 597 halos in total.
}
The center of each halo is defined by 
the position of the particle located 
at the potential minimum.
We then find self-bound, locally overdense regions in FOF groups
by {\tt SUBFIND}
\citep{2001MNRAS.328..726S}.
For the subhalo catalog,
the minimum number of particles is set to be 30.
This choice corresponds to the minimum subhalo mass
of $\sim 10^{10} \ h^{-1}M_{\odot}$.
\MS{Subhalos with mass $\sim10^{10} \ h^{-1}M_{\odot}$
can be robustly located in our simulation, but our mass resolution may not be sufficient
to estimate the density profile of the smallest subhalos. 
For stacking analysis, we use all subhalos with 30 particles or more.
We expect that our analysis is unaffected by those small subhalos,
since the HSSC signal is mostly contributed by most massive substructures
with mass of $\sim10^{12} \ h^{-1}M_{\odot}$
in group and cluster halos (see Section~\ref{subsec:halo_prop}).} 

We define the shape of a halo by applying the method 
of \citet{2006MNRAS.367.1781A}.
We assume that a halo is well described  by an ellipsoid.
The direction of the major axis and the lengths of the semimajor axes, 
$a\le b\le c$, are estimated with 
the weighted inertia moment $\tilde{I}_{ij}$:		
\beqa
\tilde{I}_{ij} &=& \sum_{n}\frac{x_{i, n}x_{j, n}}{R_{n}^2}, \\
R_{n}^2 &=& x_{n}^2/s^2 + y_{n}^2/q^2 + z_{n}^2
\eeqa
where $s=a/c$, $q=b/c$, and
$R_{n}$ is the elliptical distance in the eigenvector coordinate system from
the centre to the $n$-th particle.
The eigenvalues of $\tilde{I}$ determine the axis ratio
and the eigenvectors specify the orientation.
We estimate the tensor of $\tilde{I}$ by the iterative approach 
of \citet{2006MNRAS.367.1781A}.
The analysis begins with a sphere of $R=0.3R_{\rm vir}$
($R_{\rm vir}$ is the virial radius) and keeps the largest axis fixed at this radius. 
We remove the member particles of subhalos when we calculate $\tilde{I}$.
In summary, a halo catalog contains the information 
of $M_{200m}$,
the weighted inertia tensor $\tilde{I}$,
the positions and the masses of subhalos.

\subsection{Mock shear maps}
\label{subsec:mockmap}
We create projected mass density maps of each halo viewed along 
\MS{one axis of the orthogonal coordinate in our $N$-body simulation.}
We first generate the projected mass density map on $512^{2}$ 
two-dimensional uniform meshes by projecting
the member particles of each halo. 
The mesh size is comoving $0.025 \, h^{-1}{\rm Mpc}$.
We then derive the convergence field using Eq.~(\ref{eq:Sigma2kappa}).
\MS{
For a given convergence field, we can derive the corresponding shear field through its Fourier-space counterpart
\beqa
\tilde{\gamma}_{1}({\bd \ell}) &=& 
\frac{\ell_{x}^2-\ell_{y}^2}{\ell^2} \tilde{\kappa}({\bd \ell}),
\label{eq:kappa2shear1} \\
\tilde{\gamma}_{2}({\bd \ell}) &=& 
\frac{2\ell_{x}\ell_{y}}{\ell^2} \tilde{\kappa}({\bd \ell}),
\label{eq:kappa2shear2}
\eeqa
where quantities with tilde symbol denote the Fourier-transformed field, and ${\bd \ell}=(\ell_x, \ell_y)$ 
is the Fourier vector in the two-dimensional coordinates.
}
Note that the above procedure ignores the two effects associated with 
(1) the evolution of large-scale structure and (2) the angular size variation 
with redshift. Incorporting these effects in a direct manner
requires high-resolution ray-tracing simulations with costly ray-tracing 
\citep[e.g.,][]{2000ApJ...537....1W}. We expect the effects
are unimportant for our study on the local structure of dark matter halos.
The effect of projection of foreground large-scale structures 
is examined in detail in Section~\ref{subsec:proj_effect}.

We use a few different projected mass maps 
depending on our respective objective.
To generate a realistic lensing map around a halo,
we use all the FOF member particle.
This projected map is regarded as our fiducial one.
To study the impact of the presence of subhalos,
we generate another map by using all the particles of the main halo
(the smooth component) but removing the member particles of subhalos
whose mass $M_{\rm sub}$ is greater than
$f_{\rm sub, cut} \times M_{\rm 200m}$ where
$M_{200m}$ denotes the total halo mass.
We consider four cases 
with $f_{\rm sub, cut}=0.1, 0.05, 0.01$ and 0.
Note that $f_{\rm sub, cut}=0$ corresponds to the case 
without subhalos (removing all the subhalo member particles).
Table~\ref{tb:projection} summarizes the projected mass maps 
used in this paper.

\begin{table*}
\caption{
        Properties of five mass maps.
	\label{tb:projection}
	}
\begin{tabular}{@{}lcccl}
\hline
\hline
Map & Subhalo selection & Features \\ \hline
Fiducial & All & Fiducial model of surface mass density\\ 
Smooth & None & 
A smooth, triaxial model of surface mass density \\
Sub0.01 & $M_{\rm sub}/M_{200m}<0.01$ & 
Includes only small subhalos \\
Sub0.05 & $M_{\rm sub}/M_{200m}<0.05$ & 
Includes subhalos with the indicated mass \\
Sub0.10 & $M_{\rm sub}/M_{200m}<0.10$ & 
Includes subhalos with the indicated mass \\
\hline
\end{tabular}
\end{table*}

\section{RESULTS}
\label{sec:res}
\subsection{Dependence of halo properties}
\label{subsec:halo_prop}

We show the results of the HSSC
for our simulated halos at $z=0.33$.
We measure the three-point correlations $\zeta_{h++}$
and $\zeta_{h\kappa\kappa}$ for each halo with 
$M_{\rm 200m}\ge10^{13.5}\, h^{-1}M_{\odot}$.
When measuring the three-point correlations, 
we perform linear binning in $\theta$ 
between $\theta_{1}$ and $\theta_{2}$ with the bin width of 
0.2 times $\theta_{\rm vir}$, where $\theta_{\rm vir}$ is the corresponding angular virial radius for each halo.
We also perform linear binning in $\cos \phi$ with 
the width of 0.1.
After measuring the correlations for each halo,
we stack the measured signals over a sample of halos.
In the following, we consider the two host halo masses
with $\log (M_{200m}/h^{-1}M_{\odot})=13.5-14$ and $\log (M_{200m}/h^{-1}M_{\odot}) \ge14$.
The former contains 530 halos, while the latter contains 67 halos.

\begin{figure*}
\centering
\includegraphics[width=0.40\columnwidth, bb=0 0 547 512]
{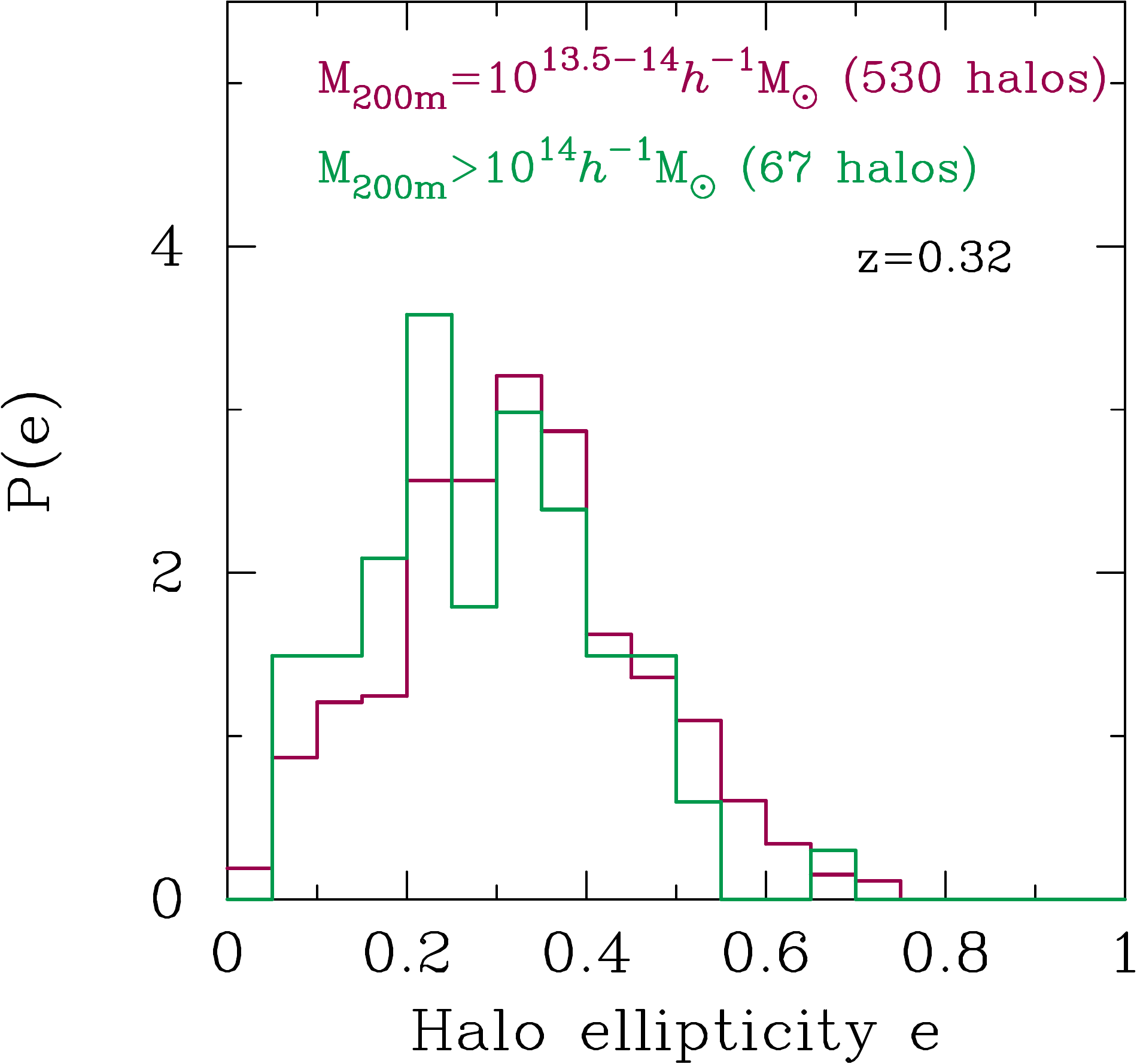}
\caption{
  The distribution of halo ellipticity in our simulation.
  Here we consider two samples: the halos with the mass of 
  $\log (M_{200m}/h^{-1}M_{\odot})=13.5-14$
  and $\log (M_{200m}/h^{-1}M_{\odot})>14$.
	}
\label{fig:P_e_sim}
\end{figure*} 

\subsubsection{Halo ellipticity}

To examine the dependence of asphericity of halos on the
HSSC, we further divide the halo sample into
two subsamples with $e \le \bar{e}$ and $e>\bar{e}$.
Here the two-dimensional ellipticity $e$ is defined by $1-q_2/q_1$
where $q_1$ and $q_2$ are the major and minor axis lengths,
and $\bar{e}$ represents the average ellipticity over the halo sample.
We evaluate $q_1$ and $q_2$ for each halo
in the same way as \citet{2003ApJ...599....7O}
by using the three-dimensional information of inertia tensor for each halo.
We find $\bar{e}=0.32\pm0.006$ and 
$0.29\pm0.015$ for halo catalogs with 
$\log (M_{200m}/h^{-1}M_{\odot})=13.5-14$ and 
$\log (M_{200m}/h^{-1}M_{\odot}) \ge14$, respectively
\MS{(the uncertainty represents the standard deviation around the mean)}.
\MS{
For the mass range of $\log (M_{200m}/h^{-1}M_{\odot})=13.5-14$,
lower-$e$ and higher-$e$ subsamples have the average ellipticity of $0.21\pm0.005$ and $0.43\pm0.006$, respectively.
For more massive halos with $\log (M_{200m}/h^{-1}M_{\odot}) \ge14$,
the average ellipticities in lower-$e$ and higher-$e$ subsamples are found to be $0.18\pm0.011$ and $0.40\pm0.014$.
Figure~\ref{fig:P_e_sim} summarizes the distribution of halo ellipticity in our simulated halos.}

\begin{figure*}
\centering
\includegraphics[width=0.70\columnwidth, bb=0 0 521 512]
{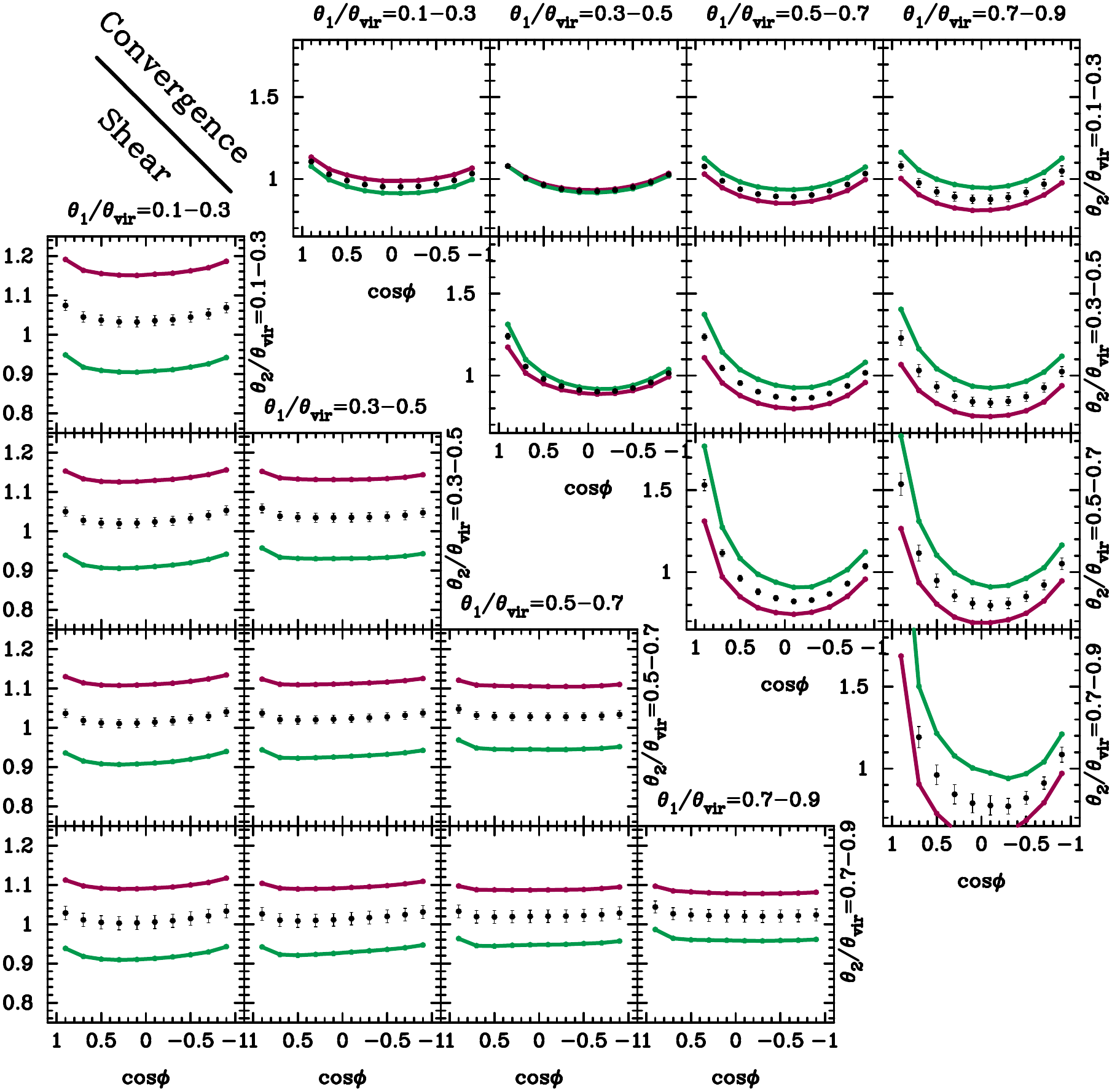}
\caption{
  The dependence of the three-point correlation functions on halo ellipticity. 
  We use the halos with the mass of 
  $\log (M_{200m}/h^{-1}M_{\odot})=13.5-14$.
  The lower-left panels show the HSSC
  $\zeta_{h++}$ as in Eq.~(\ref{eq:gggl}), 
  while the upper-right panels show 
  the HKKC, $\zeta_{h\kappa\kappa}$. 
  Note that, in each panel, the three-point correlation function is 
  normalized by the respective two-point correlations.
  \MS{The black points with error bars
  shows the results for full sample with the average ellipticity 
  of $\bar{e}=0.32$, while 
  the red and green lines are for the subsamples 
  with $\bar{e}=0.21$ and $\bar{e}=0.43$, respectively.}
  The error bars indicate the standard 1-$\sigma$ deviation.
  }
\label{fig:halo_prop_diffellip_logM135}
\end{figure*} 

\begin{figure*}
\centering
\includegraphics[width=0.70\columnwidth, bb=0 0 521 512]
{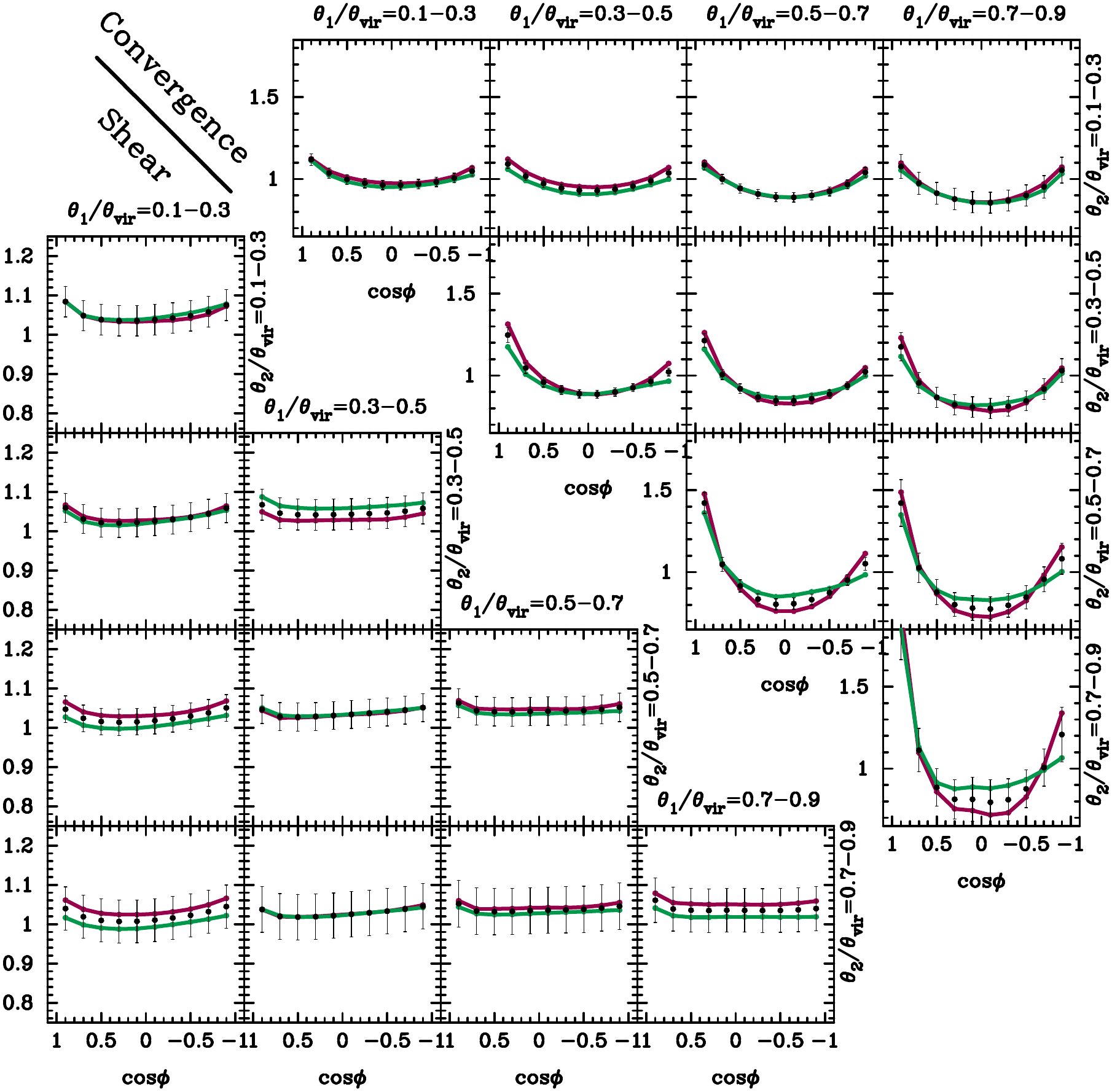}
\caption{
	As for Figure~\ref{fig:halo_prop_diffellip_logM135},
	but for more massive halos with 
	$\log (M_{200m}/h^{-1}M_{\odot})\ge14$.
    \MS{The black points with error bars
  show the results with the average ellipticity 
  of $\bar{e}=0.29$, while 
  the red and green lines are for the subsamples 
  with $\bar{e}=0.18$ and $\bar{e}=0.40$, respectively.}
	}
\label{fig:halo_prop_diffellip_logM14}
\end{figure*} 

Figures~\ref{fig:halo_prop_diffellip_logM135}
and \ref{fig:halo_prop_diffellip_logM14} show the results 
for the two-halo sample.
In each plot, the black points with error bars
show the stacked correlations without dividing the sample by 
halo ellipticity $e$, while the red and green lines are 
for the subsamples with $e\le\bar{e}$ and $e>\bar{e}$.
\MS{In the figures, panels in the lower triangular portion show the correlation
of $\zeta_{h++}(\theta_1, \theta_2, \phi)$ 
normalized by $\gamma_{t,0}(\theta_1)\gamma_{t,0}(\theta_2)$.
The upper triangular portion is for 
the three-point correlation function in terms of convergence field $\kappa$, $\zeta_{h\kappa\kappa}(\theta_1, \theta_2, \phi)$ normalized
by $\kappa_{0}(\theta_1)\kappa_{0}(\theta_2)$.}
The black error bars represent the standard deviation of
the mean signals.
Clearly, the correlation for the lower-mass halos is more sensitive to $e$ (Figure 2). 
We find that the HKKC 
at $\theta/\theta_{\rm vir}>0.3$
has a larger amplitude for halos with larger $e$.
Furthermore, HKKC tends to show a larger
amplitude at $\phi\rightarrow0$ than $\phi\rightarrow\pi$,
because massive subhalos are likely to be detected at outer region of FOF halo. The asymmetry between $\cos \phi = \pm1$ is also observed in HSSC.
We note that the signal at the innermost bin of $\theta$
will be approximated by a simple ellipsoid model presented in
\MS{Appendix~\ref{subsec:toy}.}

\subsubsection{Presence of substructures}

We next study how the presence of substructures affects 
the three-point correlations. 
In particular, we examine the mass scale of
subhalos that give significant contribution, possibly, to HSSC.
To this end, we define and use the fractional difference of the three-point
correlations 
as follows:
\beqa
\Delta \zeta (\theta_1, \theta_2, \phi; f_{\rm sub, cut})
= \frac{\zeta(\theta_1, \theta_2, \phi; f_{\rm sub, cut})}{\zeta(\theta_1, \theta_2, \phi; {\rm No\, subhalos})}-1,
\label{eq:frac_diff_zeta}
\eeqa
where 
\MS{$\zeta(\theta_1, \theta_2, \phi; {\rm No\, subhalos})$ represents
the three-point correlations without substructures and} $f_{\rm sub, cut}$ is the threshold as described in Section~\ref{subsec:mockmap}. For a given $f_{\rm sub, cut}$, we include only the subhalos whose massese are smaller than $f_{\rm sub, cut}M_{\rm 200m}$.

\begin{figure*}
\centering
\includegraphics[width=0.70\columnwidth, bb=0 0 521 512]
{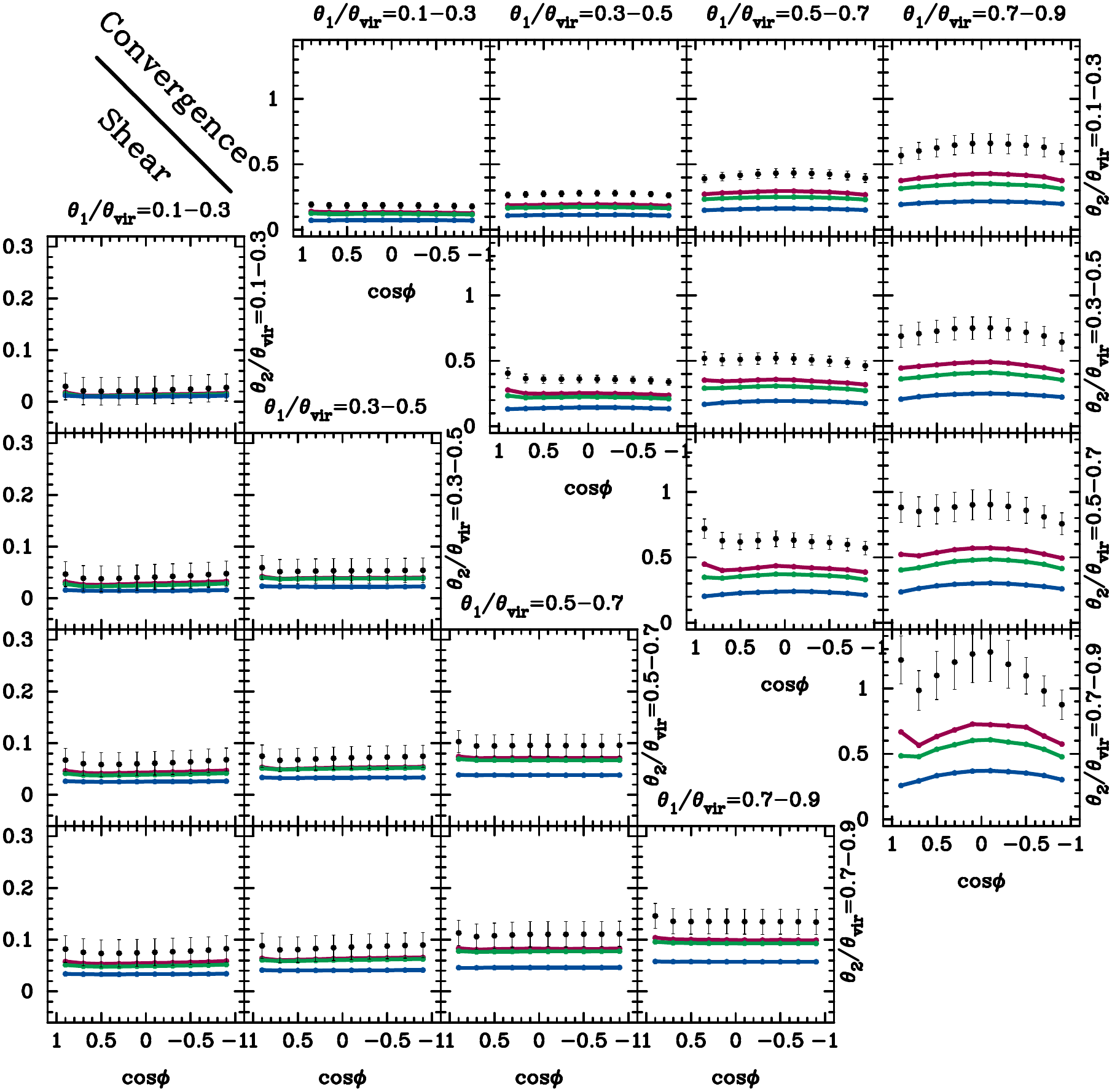}
\caption{
	The effect of subhalo abundance on three-point correlation functions.
	We plot the fractional difference with respect to the result without subhalos 
	(see Eq.~[\ref{eq:frac_diff_zeta}] for definition).
	We use halos with 
	$\log (M_{200m}/h^{-1}M_{\odot})=13.5-14$.
	The black points show the fractional difference between
	our fiducial case that includes all the subhalos
	and the result obtained after removing the subhalos. The error bars
        indicate the standard deviation evaluated in our stacking analysis.
	Different colored lines represent the results for
	the projected maps, labeled as SubXXX in 
	Table~\ref{tb:projection}.
        The red, green and blue correspond to the case of 
	Sub0.10, Sub0.05, and Sub0.01, respectively.
	Details are found in the text.
	}
\label{fig:halo_prop_diffsub_logM135}
\end{figure*} 

Figure~\ref{fig:halo_prop_diffsub_logM135} shows the result
of $\Delta \zeta$ for halos
with masses in the range of $\log (M_{200m}/h^{-1}M_{\odot})=13.5-14$.
In this figure, the black points show 
the results with $f_{\rm sub, cut}=1$ (including all the subhalos).
The red, green, blue lines represent the result 
of $f_{\rm sub, cut}=0.1$, 0.05, and 0.01, respectively.
The presence of subhalos affects more strongly the HKKC at larger $\theta$,
since more massive substructures can survive in high-density regions.
On HKKC, the contribution from subhalos 
with $M_{\rm sub}/M_{200m}\ge0.1$ can explain 
a $\sim50\%$ of amplitude over the wide range of $\theta$.
We also find that the presence of substructures 
increases the overall amplitude of $\zeta$
but does not affect the $\phi$-dependence significantly.
A typical difference due to the substructures
is found to be $\sim10\%$ for HSSC
and $\sim100\%$ for HKKC at most. 
Note that the overall trends in Figure~\ref{fig:halo_prop_diffsub_logM135}
are also found for the halo sample with $\log (M_{200m}/h^{-1}M_{\odot})>14$.

\subsection{Projection effect}
\label{subsec:proj_effect}

\begin{figure*}
\centering
\includegraphics[width=0.70\columnwidth, bb=0 0 528 512]
{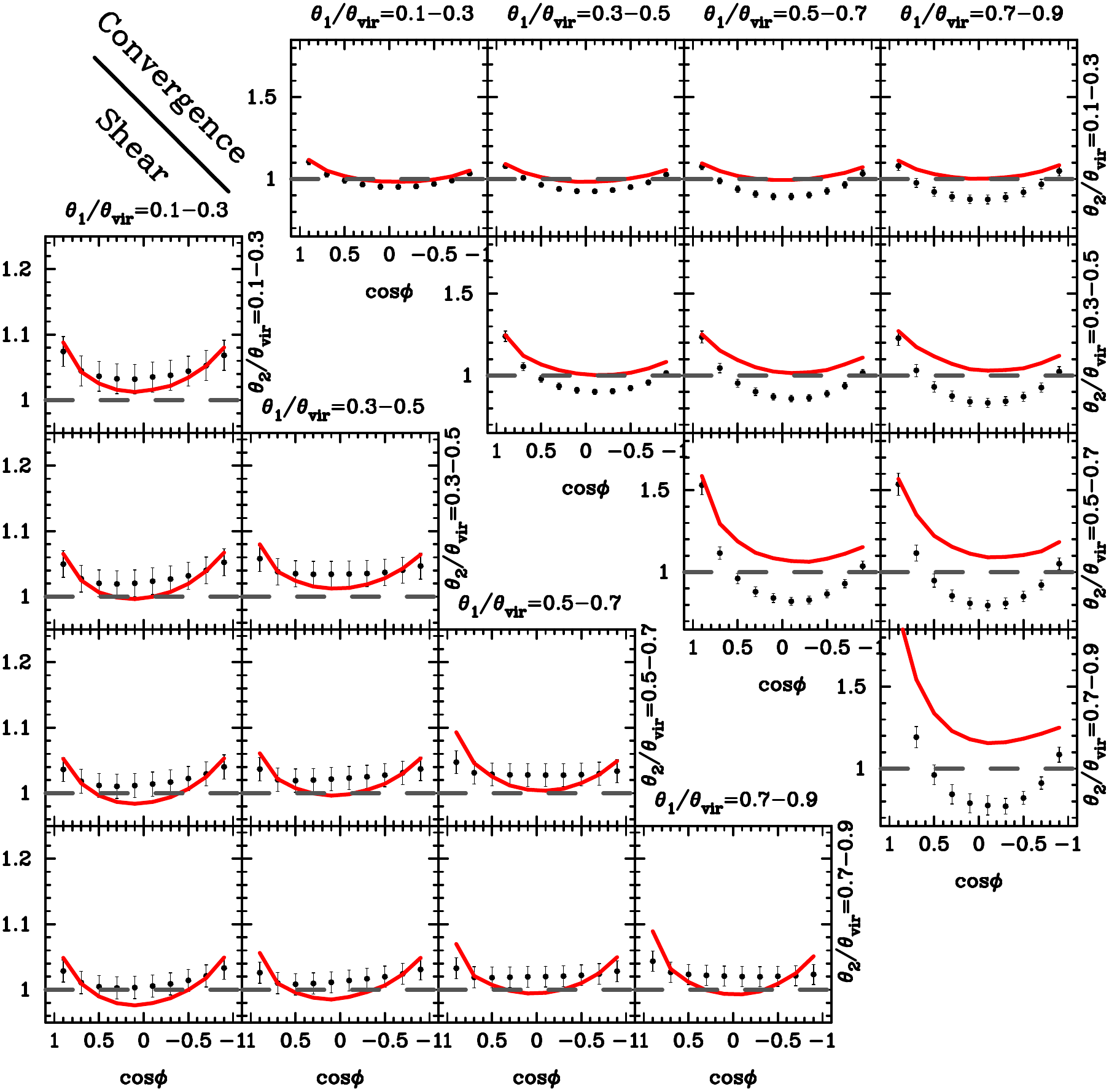}
\caption{
	The effect of projected foreground large-scale structure 
	on three-point correlation function.
	Black points with error bars show the three-point correlation
        functions for isolated halos. 
	Red lines show the expected correlation signals
	with projection effects (see also Eq.~[\ref{eq:diff_zeta_proj_est}] for detail).
        We expect that the red line and 
        the black points are close to each other
        if projection effects are dominated by uncorrelated 
        structure with the target halo.
	In this figure, we consider the halo sample with 
	$\log (M_{200m}/h^{-1}M_{\odot}) = 13.5-14$.	
	}
\label{fig:projection_effect_logM135}
\end{figure*} 

So far, we have considered the internal mass distribution of a single halo.
In practice, observed lensing signals also contain contribution from the
intervening mass distribution along a line of sight.
This contamination by uncorrelated large-scale structure is often called a
projection effect. It can possibly induce a systematic bias in the correlation
analysis. We thus study the projection effect further using our simulations
in a direct manner.
We generate the projected mass density map 
by using all the simulations particles in
a comoving box with volume of
$12.8 \times 12.8 \times L_{\rm depth} \, (h^{-1}{\rm Mpc})^3$,
where $L_{\rm depth}$ is the projection depth along the line of sight.
We set the projection depth to be $200 \, h^{-1}{\rm Mpc}$,
the box size of our N-body simulations.
Note that we fix the mesh size of 
comoving $0.025 \, h^{-1}{\rm Mpc}$ 
and the number of meshes ($=512^2$) as in Section~\ref{subsec:mockmap}.

The projection effect due to uncorrelated structure along a line of sight
can be estimated, and hence can be subtracted, by performing
the same measurement of the correlation functions around random points.
We measure the HSSC and HKKC around random points
and construct the following estimator of correlation:
\beqa
\zeta_{\rm est} (\theta_1, \theta_2, \phi) 
&=& 
\zeta(\theta_1, \theta_2, \phi; {\rm All})
-\zeta (\theta_1, \theta_2, \phi; {\rm Rand}) \nonumber \\
&&
\,\,\,\,\,\,\,\,\,\,
\,\,\,\,\,\,\,\,\,\,
\,\,\,\,\,\,\,\,\,\,
-\left[ \xi(\theta_1; {\rm All}) -\xi (\theta_1; {\rm Rand})\right]\xi (\theta_2; {\rm Rand})
-\left[ \xi(\theta_2; {\rm All}) -\xi (\theta_2; {\rm Rand})\right]\xi (\theta_1; {\rm Rand}), 
\label{eq:diff_zeta_proj_est}
\eeqa
where 
$\zeta({\rm All})$ represents the three-point correlation functions
measured including all the simulation particles
and with $\zeta({\rm Rand})$ being the measured correlation
around random points. Similarly, we define $\xi({\rm All})$ and $\xi({\rm Rand})$ 
as the two-point correlations around halos and random points, respectively.
We have tested and confirmed that the correlation functions around random points
converge when the number of random points is set to be ten times as large as
the number of halos.

Figure~\ref{fig:projection_effect_logM135} shows 
the result of $\zeta_{\rm est}$ for the halo catalog with the selection of 
$\log (M_{200m}/h^{-1}M_{\odot}) = 13.5-14$.
We find that, for HSSC, the signals from the target halos
can be recovered after the random contribution is subtracted.
On the other hand, the same technique does not work for HKKC.
HKKC directly probes the projected mass distribution
along a line of sight, and thus the contribution from neighboring
(in projection) halos
is significant even at small angular scales.
The difference between the black points and the red line manifests
the projection effect on the HSSC and HKKC;
projection of the intervening structure that is
physically close to, and hence is correlated with, the lense
halo can induce a few percent effects for the HSSC,
while it can affect the HKKC by a factor of about $\sim20\%$.
The projection effect is more prominent at large angular scales
(large $\theta$), as expected.
 
\subsection{Detectability and implication}
\label{subsec:detectability}
In this section, we explore the detectability of HSSC as 
defined in Eq.~(\ref{eq:gggl}) and discuss the implication for
probing the internal structures of galaxy-group halos
and cluster halos.

In the following statistical analysis, we use the HSSC
measured with the radial bins of $\theta \, [{\rm arcmin}] = (0.76, 2.29)$
and the azimuthal bins of $\cos \phi=(-0.8, -0.4, 0, 0.4, 0.80)$.
The bin width of $\theta$ is set to be 16 times $\theta_{\rm mesh}$,
where $\theta_{\rm mesh}=0.095\, {\rm arcmin}$ 
corresponds to the angular size of a mesh in the projected mass map.
We use 15 data values of HSSC in total.
Note that the binning in radial direction roughly 
corresponds to $\theta/\theta_{\rm vir}\simlt0.5$
for the mass-limited sample with $M>10^{14}\, h^{-1}M_{\odot}$
at $z=0.3$.
We limit the range of radius as $\theta/\theta_{\rm vir}\simlt0.5$
because the signals at the outer regions are likely affected by
the projection of the neighboring, and hence {\it correlated}, structures
(see e.g., Figure~\ref{fig:projection_effect_logM135})

\subsubsection{Total signal-to-noise ratio}

To quantify the detectability of HSSC, we calculate
the cumulative signal-to-noise ratio defined as
\beqa
\left(S/N\right)^2 = \sum_{i, j} 
\zeta_{h++}(i) {\bd C}^{-1}_{ij} \zeta_{h++}(j),
\label{eq:s2n}
\eeqa
where indices $i$ and $j$ run over the triangle configurations of interest,
$\zeta_{h++}(i)$ denotes the HSSC for the $i$-th
triangle configulation, 
{\bd C} is the covariance matrix
and ${\bd C}$ is its inverse.

We evaluate the covariance matrix of ${\bd C}$ on the assumption that
the covariance is dominated by the intrinsic ellipticity of sources, the so-called shape noise.
We model the shape noise $\epsilon$ by drawing the two-dimensional
Gaussian random field on $512^2$ mesh points:
\beqa
P(\epsilon) = \frac{1}{\pi \sigma^2} \exp \left(-\frac{\epsilon_1^2 + \epsilon_2^2}{\sigma^2}\right),
\eeqa
where $\sigma^2=\sigma_{\rm obs}^2/n_{\rm gal}\theta_{\rm mesh}^2$ with $\theta_{\rm mesh}=0.095$ arcmin, 
\MS{$\sigma_{\rm obs}$ is the rms of observed source ellipticities},
and $n_{\rm gal}$ is the number density of sources.
In this paper, we assume $\sigma_{\rm obs}=0.4$
and $n_{\rm gal}=30\, {\rm arcmin}^{-2}$,
which are expected in a future imaging survey 
\MS{
by Large Synoptic Survey Telescope
\citep[LSST; ][]{2009arXiv0912.0201L}.
}
\MS{Note that $\sigma_{\rm obs}$ here includes both
galaxy shape noise and measurement uncertainty.
\citet{2007ApJS..172..219L} have found that 
the intrinsic galaxy shape noise was found empirically to be $0.23$ per reduced shear component, while the image simulation for LSST survey
indicates that the measurement error in LSST
would be of an order of $0.1$ per shear component with r-band limiting magnitude of $\sim24$ \citep[][]{2013ApJ...774...49B}.
Hence, we expect $\sigma_{\rm obs}/\sqrt{2}\sim\sqrt{0.1^{2}+0.23^2}=0.25$ per component and this is close to our assumed value of $0.4/\sqrt{2}=0.28$
\footnote{
\MS{
It is also worth noting that the signal-to-noise ratio will scale with
$(\sigma_{\rm obs}/0.4)^{-2}$ for a given $n_{\rm gal}$.
If assuming more realistic value of $\sigma_{\rm obs}=0.365$, we expect 
the signal-to-noise ratio will be improved by a factor of $\sim1.2$.}
}.}
We then create 10,000 realizations of noise map and perform
the measurements of HSSC in Eq.~(\ref{eq:gggl})
with respect to the center of map.
Using 10,000 HSSCs for the shape noise, 
we compute the covariance matrix as 
\beqa
\bar{\zeta}_{h++, n} (i)
&=& \frac{1}{N_{r}} \sum_{r=1}^{N_{r}} 
\zeta_{h++, n}(i; r), \\
{\bd C}^{(\rm noise)}_{ij}
&=& \frac{1}{N_{r}-1}\sum_{r=1}^{N_{r}} 
\left[\zeta_{h++, n}(i; r) - \bar{\zeta}_{h++, n} (i)\right]
\left[\zeta_{h++, n}(j; r) - \bar{\zeta}_{h++, n} (j)\right],
\eeqa
where $\zeta_{h++, n}(i; r)$ represents 
the HSSC for 
the $i$-th triangle configuration and $r$-th realization of noise map,
and $N_{r} = 10000$.
Since ${\bd C}^{(\rm noise)}$ corresponds to the statistical uncertainty
\MS{for individual halos}, 
we derive ${\bd C}$ by scaling ${\bd C}^{(\rm noise)}$ 
with the number of halos $N_{\rm halo}$:
\beqa
{\bd C} = \frac{1}{N_{\rm halo}}{\bd C}^{(\rm noise)}.
\label{eq:cov_sn}
\eeqa

Using Eqs.~(\ref{eq:s2n}) and (\ref{eq:cov_sn}),
we compute the expected $S/N$ for the mass-limited halos 
at $z=0.33$ as
\beqa
S/N &=& 1.01 \, \sqrt{N_{\rm halo}} \,\,\,\, (M_{200m}\ge10^{13.5}h^{-1}M_{\odot}), \\
S/N &=& 2.18 \, \sqrt{N_{\rm halo}} \,\,\,\, (M_{200m}\ge10^{14}h^{-1}M_{\odot}),
\eeqa
where we use the stacked signals for simulated halos in computing
of $S/N$.
Hence we expect the high-significance measurement of the HSSC 
using data from future galaxy imaging surveys.
We will be able to achieve 5\% measurement of HSSCs
by using $\sim$400 lenses on galaxy-group scales
and $\sim$100 lenses on galaxy-cluster scales.

\begin{table*}
\caption{
	Our model specifications with different properties 
	of the structure of dark matter halos.
	We perform the subdivision presented in this table 
	for the mass-limited sample
	with $M_{\rm 200m}\ge10^{13.5}\, h^{-1}M_{\odot}$
	and $10^{14}\, h^{-1}M_{\odot}$.
	\label{tb:diff_models}
	}
\begin{tabular}{@{}lccccl}
\hline
\hline
Model & Halo ellipticity & Subhalo selection & Features \\ \hline
Fiducial & $e=0-1$ & All & Our fiducial model with average $e = 0.3$\\ 
\hline
Spherical & $e=0$ & None & $\zeta_{h++}(\theta_1, \theta_2, \phi) = \xi_{h+}(\theta_1)\xi_{h+}(\theta_2)$ \\
\hline
Low-$e$ & $e=0-0.3$ & All & averaged $e = 0.2$ \\
Mid-$e$ & $e=0.3-0.6$ & All & averaged $e = 0.4$ \\
High-$e$ & $e=0.6-0.9$ & All & averaged $e = 0.65$ \\
\hline
Smooth & $e=0-1$& None & No subhalos \\
Sub0.05 & $e=0-1$& $M_{\rm sub}/M_{200m}<0.05$ & 
Includes only small subhalos \\
Sub0.10 & $e=0-1$& $M_{\rm sub}/M_{200m}<0.10$ & 
Includes massive subhalos \\
\hline
\end{tabular}
\end{table*}

\subsubsection{Inferring halo properties}

We next examine the ability of the HSSC
as a probe of the internal structure
of dark matter halos.
We define and use the following quantity as a measure of
information content:
\beqa
\Delta \chi^2 = \sum_{ij} 
\left[\zeta_{h++}(i; {\rm test})-\zeta_{h++}(i; {\rm fid})\right]
{\bd C}_{ij}^{-1}
\left[\zeta_{h++}(j; {\rm test})-\zeta_{h++}(j; {\rm fid})\right],
\label{eq:deltachi2}
\eeqa
where $\zeta_{h++}(i; {\rm test})$
represents the model HSSC
at the $i$-th bin (that we aim at detecting),
$\zeta_{h++}(i; {\rm fid})$
is the HSSC of the fiducial model,
and 
${\bd C}$ is the covariance defined by Eq.~(\ref{eq:cov_sn}).
\MS{We evaluate the covariance assuming 
a future imaging survey like LSST.}
In this section, we define both of 
$\zeta_{h++}(\rm test)$ and $\zeta_{h++}(\rm fid)$
by using our simulated halos and projected mass density maps.
For our fiducial model, we measure the HSSCs
for a mass-limited sample with $M_{\rm 200m}\ge M_{\rm thre}$,
with two cases of 
$M_{\rm thre}=10^{13.5}\, h^{-1}M_{\odot}$
and 
$M_{\rm thre}=10^{14}\, h^{-1}M_{\odot}$.
We expect that the former sample corresponds to
massive galaxies in Sloan Digital Sky Survey 
\citep[e.g.][]{2001AJ....122.2267E},
while the latter is for the sample of galaxy clusters 
identified in optical imaging surveys \citep[e.g.][]{2014ApJ...785..104R}.

For a given $M_{\rm thre}$, 
the simplest reference model assumes
spherical halos and no subhalos.
Then the expected correlation is expressed as
$\zeta_{h++}(\theta_1, \theta_2, \phi) = \xi_{h+}(\theta_1)\xi_{h+}(\theta_2)$
where 
$\xi_{h+}$ is defined in Eq~(\ref{eq:ggl}).
Also there is no azimuthal dependence in the correlation.
We investigate the sensitivity on 
halo ellipticity $e$ and the abundance of subhalos.
To this end, we divide the mass-limited sample 
into three subsamples by halo ellipticity 
or the abundance of subhalos.
We use three ellipticity bins: $e=0-0.3, 0.3-0.6$, and $0.6-0.9$.
Note that we include all the resolved subhalos 
when calculating the HSSC for these subsamples.
Similarly, we consider the dependence of subhalo abundance on
the HSSC by using different projected mass maps
as in Table~\ref{tb:projection}.
We measure the HSSCs by using maps 
named as ``Smooth", ``Sub0.05", and ``Sub0.10".
The first one corresponds to the model in the absence of subhalos,
while the last two are for the model including 
the subhalo with some cutoff of subhalo masses.
Table~\ref{tb:diff_models} summarizes our subdivision of mass-limited samples.

\begin{table*}
\caption{
	The expected statistical power of the HSSC to
	constrain the internal structure of dark matter halos.
	We here present $\Delta \chi^2$ defined 
	by Eq.~(\ref{eq:deltachi2}).
	When computing $\Delta \chi^2$, 
	we refer the models summarized in
	Table~\ref{tb:diff_models}.
	\label{tb:deltachi2}
	}
\scalebox{0.85}[0.85]{
\begin{tabular}{@{}lccccccccl}
\hline
\hline
$M_{\rm thre}\, [h^{-1}M_{\odot}]$ 
& Spherical & Low-$e$ & Mid-$e$ & High-$e$ 
& Smooth & Sub0.05 & Sub0.10 \\ \hline
$10^{13.5}$
& $4.14\, (N_{\rm halo}/100)$ 
& $1.72\, (N_{\rm halo}/100)$ 
& $0.816\, (N_{\rm halo}/100)$ 
& $13.6\, (N_{\rm halo}/100)$ 
& $9.83\, (N_{\rm halo}/10^4)$ 
& $1.78\, (N_{\rm halo}/10^4)$ 
& $1.21\, (N_{\rm halo}/10^4)$ \\
\hline
$10^{14}$
& $25.6\, (N_{\rm halo}/100)$ 
& $0.0938\, (N_{\rm halo}/100)$ 
& $0.251\, (N_{\rm halo}/100)$ 
& $90.7\, (N_{\rm halo}/100)$ 
& $78.7\, (N_{\rm halo}/10^4)$ 
& $15.4\, (N_{\rm halo}/10^4)$ 
& $6.01\, (N_{\rm halo}/10^4)$ \\
\hline
\end{tabular}
}
\end{table*}

We summarize the result of $\Delta \chi^2$
for the mass-limited sample with various models of 
halo ellipticity and subhalo abundance in Table~\ref{tb:deltachi2}.
\MS{The results in the table are obtained assuming the expected data quality in an LSST-like future imaging survey.}

First of all, the HSSC is found to be 
powerful for examining the simplest model of halo structure as spherical smoothed distribution.
For $M_{\rm thre}=10^{14}\, h^{-1}M_{\odot}$,
we find $\Delta \chi^2 \simeq26$ with 100 halos,
showing the spherical halos can be 
rejected at $\sqrt{26}\simeq5\sigma$ significance level.
Considering massive galaxies with 
$M_{\rm thre}=10^{13.5}\, h^{-1}M_{\odot}$,
we will be able to distinguish between 
a perfectly spherical and smooth model from our fiducial model
at the $5\sigma$ significance level
when applying the HSSC analysis to 1000 halos.

Moreover, if we can use 100 halos, 
we will be able to obtain $\Delta \chi^2 \simeq 14$ and $90$
for the high-$e$ sample 
with $M_{\rm thre}=10^{13.5}\, h^{-1}M_{\odot}$
and $M_{\rm thre}=10^{14}\, h^{-1}M_{\odot}$, respectively.
Note that the average elliptically $\bar{e}$ is found to be 0.65
for both $M_{\rm thre}$ when we use the high-$e$ sample,
while $\bar{e}\simeq0.3$ is found for our fiducial cases.
Hence, using 100 lenses, 
we can distinguish the model of $\bar{e}\simeq0.3$
with $\bar{e}=0.65$ with the significance level of 
$\sqrt{14}\simeq 3.7$ for massive galaxies,
while the significance will be $\sqrt{90}\simeq9.5$ 
for galaxy clusters.
On the other hand, we need a larger number of halos
in order to constrain 
the average halo ellipticity with the level of 0.1.
Table~\ref{tb:deltachi2} shows $\sim1000$ halos
are necessary to detect $3\sigma$ difference
between $\bar{e}=0.3$ and $\bar{e}=0.2$ or $0.4$
for $M_{\rm thre}=10^{13.5}\, h^{-1}M_{\odot}$.
On the cluster-sized halos, we find the small difference of HSSCs
between $\bar{e}=0.3\pm0.1$, showing much more ($\sim10,000$)
halos are required
to improve the constraints of $\bar{e}$.
It is worth noting that our definition of $e$ is roughly 
based on the region within 0.3 times virial radius 
and the expected constraint of $e$ will depend on its definition.

To infer the abundance of subhalos, we typically require $\sim10^4$ halos
to obtain $\Delta\chi^2 > 1$ for either $M_{\rm thre}$.
If we consider 10,000 halos with 
$M_{\rm thre}=10^{13.5}\, h^{-1}M_{\odot}$,
we will be able to contain the model in the absence of subhalo
with $\sim3\sigma$ significance,
whereas $\sim10^5$ halos are needed to constrain 
the mass function of subhalos.
For the cluster-sized halo, $\sim1500$ halos are sufficient 
to reject the model in the absence of subhalo with $\sim3\sigma$ significance, while we need $\sim10^4$ halos 
to constrain the subhalo mass function.
Therefore, to constrain both of halo ellipticity and 
subhalo mass function, we require $10^5$ objects 
for massive galaxies and $10^4$ objects for clusters.
\MS{Note that the large number of massive galaxies and clusters 
are already available in the current galaxy survey
\citep[e.g.][]{2014MNRAS.441...24A, 2014ApJ...785..104R},
but we need a deeper imaging survey to collect a large number
of source galaxies and to measure their shapes.}




In the above, we assume that the position of each halo center
is precisely known.
In real observations, the central position of a
cluster-sized halo is assumed to be the position of the brightest cluster galaxy (BCG).
Recent observations show that
the positions of BCGs are distributed around the
the projected halo centers \citep[e.g.,][]{2010MNRAS.405.2215O, 2012MNRAS.426.2944Z}.
This off-centering effect has been studied for massive galaxies in \citet{2013MNRAS.435.2345H}.
In the Appendix, we present a simple model of off-centering effect
on the three-point correlations. 
There, we show that the off-centering typically reduces the value of $\Delta \chi^2$
in Table~\ref{tb:deltachi2} by a factor of $\sim0.5-0.6$.
However, the HSSC can 
still constrain the internal halo structures on a statistical basis.

\section{CONCLUSION AND DISCUSSION}
\label{sec:con}

We have studied the three-point correlation 
of the distribution of dark matter halos and
the tangential shears of two background sources,
referred as to halo-shear-shear correlation (HSSC).
We have used the outputs of high-resolution cosmological $N$-body simulations of
the standard $\Lambda$CDM model to
generate realistic projected density fields around dark halos.
Finally, we have studied the information content of HSSCs
and quantified the statistical significance of measurement of
halo ellipticity and subhalo abundance.
Our findings are summarized as follows:

\begin{enumerate}


\vspace{2mm}
\item
HSSC is sensitive to the halo ellipticity $e$
within the virialized region.
However, 
the three-point correlation functions measured for halos in
our cosmological simulations are not fully explained by
an elliptical halo model.
This is because massive subhalos 
are populated in the outer region of the halo, generating
large correlations.
HKKC can have an asymmetry between $\phi\rightarrow0$
and $\phi\rightarrow\pi$ even if all the resolved subhalos 
are removed.
This suggests that even the smooth component of the
density field within the halo's virial radius
is different from that of a simple ellipsoid.
Halo ellipticity can contribute to the HSSC normalized
by its angle average about $\sim5-10\%$
in the wide range of radii.
The sensitivity of the HSSC on $e$ depends on the halo mass.

\vspace{2mm}
\item
The presence of subhalos can affect the HSSC
significantly at larger radii.
Subhalos with mass greater than 10\% of the host halo mass
can give a $\sim50\%$ contribution to the amplitude of the HKKCs.
Also, subhalos do not cause significant
azimuthal variation of the three-point correlations. 
Interestingly, the fractional contribution from subhalos to the HSSC 
is roughly independent of the host halo mass.

\vspace{2mm}
\item
\MS{We study the detectability of the HSSC assuming 
the specifics of LSST.}
\MS{When assuming source number density of $30\, {\rm arcmin}^{-2}$,
the observed scatter of shape of $0.4$,
and source redshift of 1,
we expect signal-to-noise (S/N) ratio of the HSSC
is $\sim20$ with $\sim400$ massive galaxies.}
A similar S/N can be obtained with $\sim100$ galaxy clusters.
A spherical smoothed mass model can be ruled out with $\sim5\sigma$
significance level by the HSSC when $100-1000$ halos are observed.
For 1000 galaxies with mass greater than 
$10^{13.5}\, h^{-1}M_{\odot}$, 
we can distinguish $\bar{e}=0.3$
and $\bar{e}=0.3\pm0.1$ with a $3\sigma$ significance.
On the cluster-sized halos, we need $\sim$10,000 lenses
to distinguish $\bar{e}=0.3$ and $\bar{e}=0.3\pm0.1$
at $3\sigma$ significance.
To constrain the mass function of subhalos, 
we need $\sim10^5$ and $\sim10^4$ halos for massive galaxies
and clusters, respectively.
Note that observation of such a large number of galaxies
and galaxy clusters is indeed possible with
the current-generation galaxy surveys
\citep[e.g.][]{2014MNRAS.441...24A, 2014ApJ...785..104R},
\MS{but our proposed statistics 
require wide and deep imaging data of background galaxies in the future.}

\end{enumerate}

Future studies should focus on studying possible systematic effects on
the measurement of HSSC.
There are likely uncertainties or measurement errors associated
with the off-centering effect,
the dilution effect by satellite galaxies,
the intrinsic alignment of satellite galaxies,
and the baryonic effect on the underlying mass distribution
\citep[][]{2015ApJ...806..186O}.
It is also necessary to develop an accurate theoretical 
model of the shape and the internal structure of dark matter halos for non-standard dark matter models
such as self-interacting dark matter.
Future imaging surveys will provide us with rich data
that allow us to perform statistical analyses to reveal the
nature of dark matter.

\section*{acknowledgments}
We thank Masamune Oguri for useful discussions and comments on the manuscript.
M.S. is supported by Research Fellowships of the Japan Society for the Promotion of Science (JSPS) for Young Scientists.
N.Y. and  M.S. acknowledge financial support from JST CREST (JPMHCR1414).
Numerical computations presented in this paper were in part carried out
on the general-purpose PC farm at Center for Computational Astrophysics,
CfCA, of National Astronomical Observatory of Japan.

\bibliographystyle{mnras}
\bibliography{bibtex}

\appendix
\section{Toy model}
\label{subsec:toy}
Here we present simple two models of 
\MS{surface mass density at lens plane}
and use them to derive the three-point correlation of
Eq.~(\ref{eq:gggl}). One assumes the ellipsoid surface mass density and 
another considers the case of spherical surface mass density 
in the presence of substructures.
\MS{Although these simple models will not be suitable for predicting actual HSSC measurements, they are still helpful to understand to what extent the HSSC contains meaningful information of shape and substructures of dark matter halos.}

\begin{figure*}
\centering
\includegraphics[width=0.70\columnwidth, bb=0 0 521 512]
{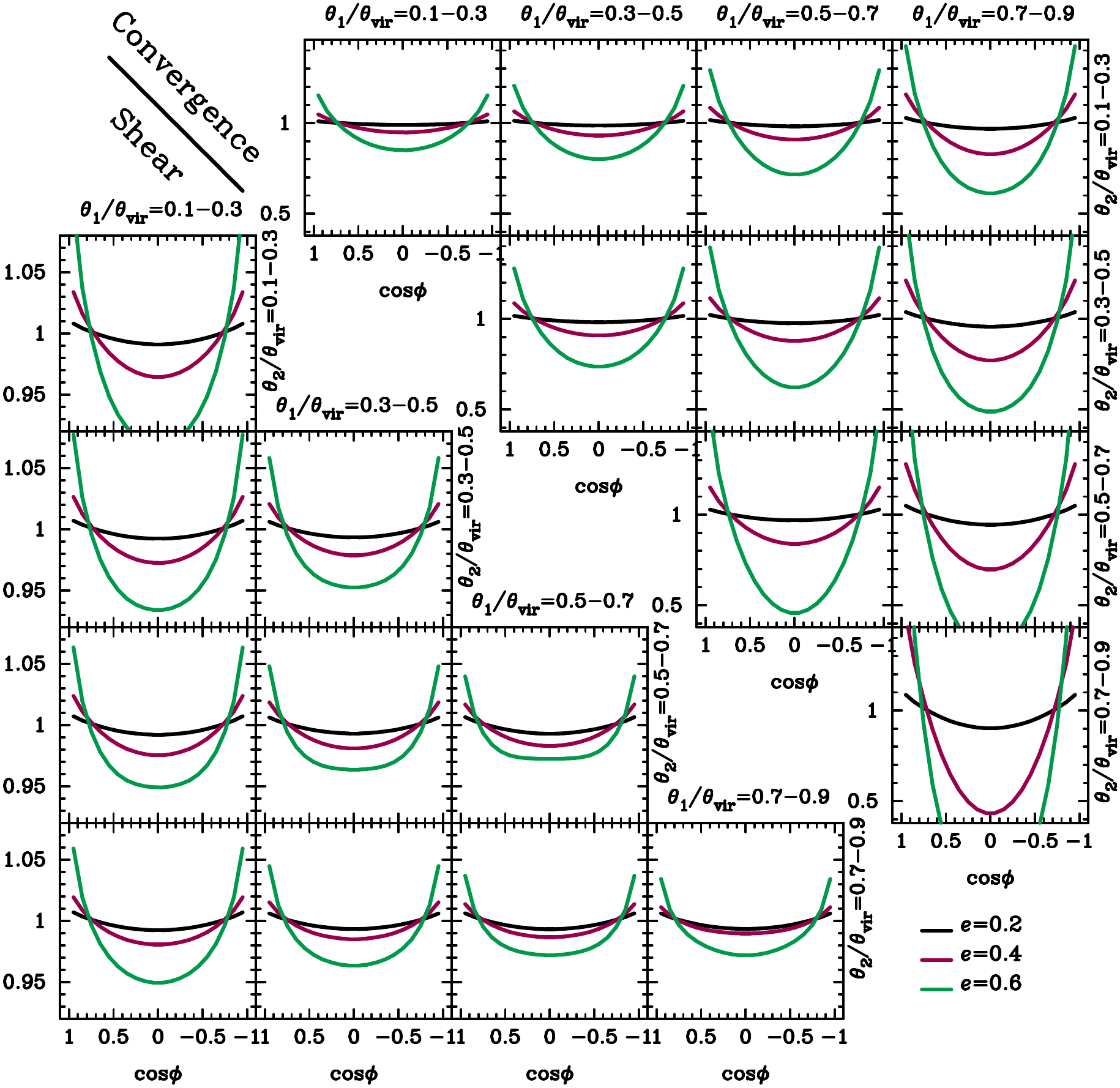}
\caption{
	The three-point correlation functions for elliptical NFW halos 
	with $M=10^{14}\, h^{-1}M_{\odot}$, 
	$z=0.3$ and $c_{\rm NFW}=5.6$.
	Panels in the lower-left portion show the correlation for tangential shear
	$\zeta_{h++}$ (Eq.~[\ref{eq:gggl}]), 
	while the upper-right panels show
	the results for convergence, $\zeta_{h\kappa\kappa}$. 
	Note that the three-point correlation function is normalized by
    the respective halo-shear or halo-convergence two-point correlations at given $\theta_{1}$ and $\theta_{2}$.  
	The black, red, green, and blue curves in each panel 
	are the results with halo ellipticity $e=0.2$, 
	0.4, and 0.6, respectively. 
        In each panel, we plot the correlation function measured 
	in the radial bin between $\theta_{1}$ and $\theta_{2}$.
	The radial coordinate $\theta$ 
	is given in units of the angular virial radius of the halo.
	}
\label{fig:3pcf_ellip_halo}
\end{figure*} 

\subsection{Elliptical halo}
\label{subsubsec:ellip_halo_model}
It is well known that the density profile of CDM halos can be approximated as
a sequence of the concentric triaxial distribution
\citep[e.g.][]{2002ApJ...574..538J}.
As a reference, we adopt the following mass model:
\beqa
\kappa(\theta, \varphi)
&=& \kappa_{\rm sph}(\vartheta), \label{eq:kappa_ellip} \\ 
\vartheta^2 
&=& \theta^2 \left[
(1-e)^{-1}\cos^2 \varphi+(1-e)\sin^2 \varphi
\right],
\eeqa
where
\MS{$\theta$ and $\varphi$ are the radial 
and azimuthal angles in polar coordinate
(the origin is set be the halo centre),}
$\kappa_{\rm sph}$ represents 
the radial convergence profile for the spherical NFW profile
\citep[e.g.][]{1996A&A...313..697B}.
The spherical NFW profile can be specified 
by two parameters, the halo concentration $c_{\rm NFW}$
and mass $M$.
Here we define the mass 
$M=4\pi r^3_{\rm vir} \Delta_{\rm vir}(z) \rho_{\rm crit}(z)/3$, 
where $\rho_{\rm crit}(z)$ is the critical 
density at cluster redshift $z$,
$r_{\rm vir}$ is the virial radius corresponding to the overdensity 
criterion $\Delta_{\rm vir}(z)$ (as shown in, e.g., \citet{1997ApJ...490..493N}).
In Eq.~(\ref{eq:kappa_ellip}), the halo ellipticity $e$ 
is defined as $1-q_2/q_1$ where $q_1$ and $q_2$ are
the major and minor axis lengths of the iso-density contour, 
respectively.
Note that the model of Eq.~(\ref{eq:kappa_ellip}) is consistent 
with recent observations of galaxy cluster halos
\citep{2010MNRAS.405.2215O, 2012MNRAS.420.3213O}.
For the mass model in Eq.~(\ref{eq:kappa_ellip}), 
we first generate the convergence field in grids, 
Fourier-transform the convergence field to obtain the Fourier coefficients of the shear field at each grid 
\MS{(see Eqs.~(\ref{eq:kappa2shear1}) and (\ref{eq:kappa2shear2}))}, 
and then compute the grid-based shear field 
from the inverse Fourier transform.

Figure~\ref{fig:3pcf_ellip_halo} shows 
the HSSC for 
an elliptical halo.
Here we set $M=10^{14}\, h^{-1}M_{\odot}$,
$z=0.3$, and $c_{\rm NFW}=5.6$, 
but vary the halo ellipticity $e$.
The lower triangular panels show the correlation
of $\zeta_{h++}(\theta_1, \theta_2, \phi)$ 
normalized by $\gamma_{t,0}(\theta_1)\gamma_{t,0}(\theta_2)$.
Similarly, the upper triangular panels are for 
the three-point correlation function
of $\zeta_{h\kappa\kappa}(\theta_1, \theta_2, \phi)$ 
normalized by $\kappa_{0}(\theta_1)\kappa_{0}(\theta_2)$.
Figure~\ref{fig:3pcf_ellip_halo} shows that
the normalized three-point functions for an elliptical halo have similar
amplitudes at different radii, but
stronger azimuthal dependence for larger halo ellipticity. 
Moreover, the correlation 
for a squeezed triangle configuration of $\phi \rightarrow 0$ or $\pi$ 
has larger amplitudes, since the mass density increases along the major
axis as in Eq.~(\ref{eq:kappa_ellip}).
In general, HSSC shows a smaller azimuthal dependence than 
HKKC because of the non-local nature of the shear field.
In other words, the shear field contains generically information on the surface mass densities
over a relatively wide region around the line of sight (e.g., see Eq.~[\ref{eq:she_t_0}]).
This fact makes it difficult to observe fine features in the local surface mass density
with HSSC.

\subsection{Spherical halo with substructures}

\begin{figure*}
\centering
\includegraphics[width=0.70\columnwidth, bb=0 0 521 512]
{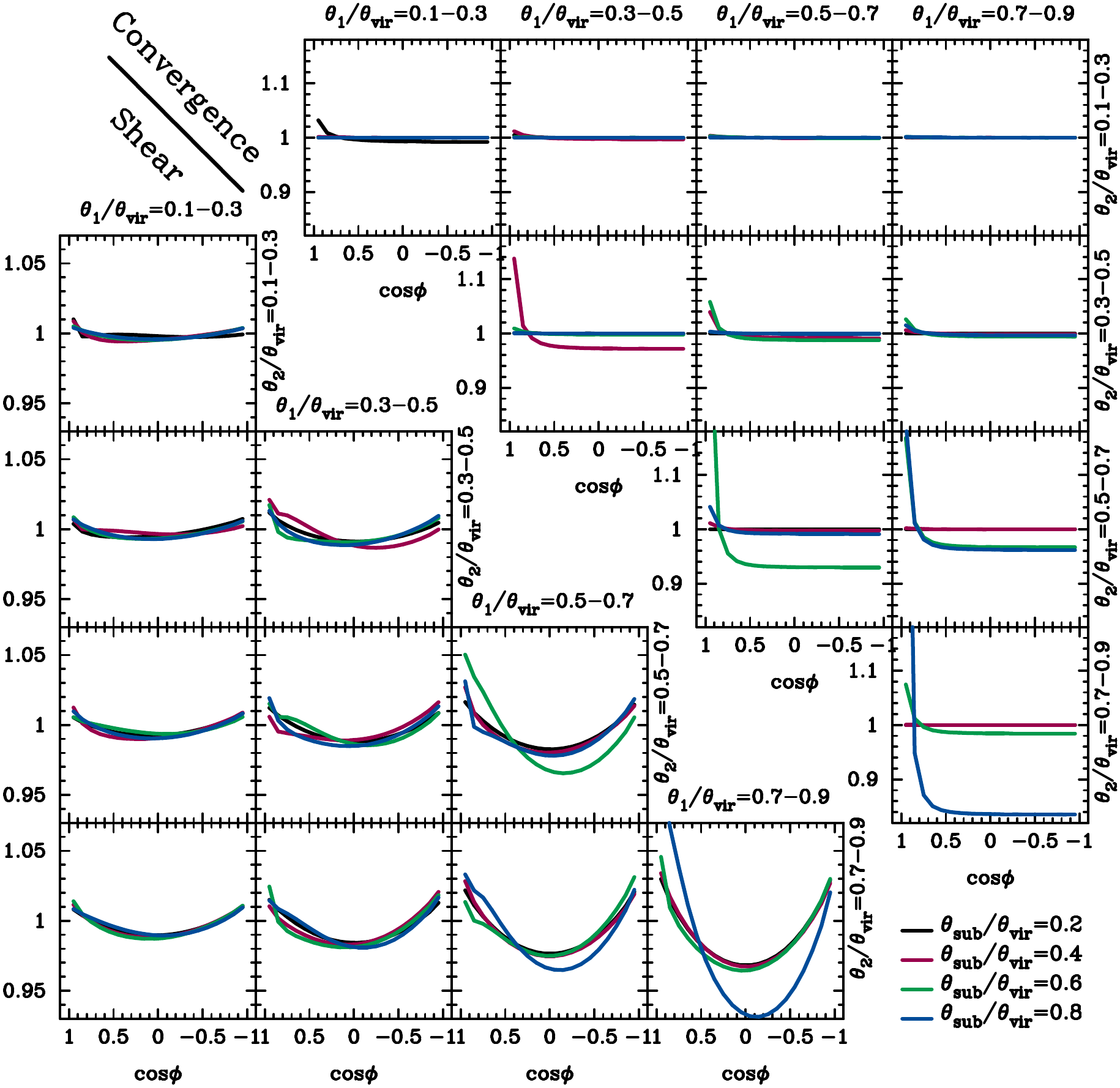}
\caption{
  We plot $\zeta_{h++}$ and $\zeta_{h\kappa\kappa}$ as for Figure~\ref{fig:3pcf_ellip_halo},
  but with the effect of substructures.
  Here, we consider a mass model that assumes
  a spherical NFW main halo with a truncated NFW subhalo.
  We set 
  the main halo mass to be 
  $M_{\rm host}=10^{14}\, h^{-1}M_{\odot}$,
	the concentration $c_{\rm NFW, host}=5.6$,
	and the subshalo mass
	$M_{\rm sub}=10^{13}\, h^{-1}M_{\odot}$,
	and the subhalo concentration $c_{\rm sub}=7$.
	The black, red, green and blue lines in each panel
	correspond to the model with the subhalo position at
	$\theta_{\rm sub}/\theta_{\rm vir}=0.2, 0.4, 0.6$ and 0.8,
	respectively.	
	We generate 300 realizations for each model, 
	and then calculate the ensemble-averaged three-point functions. 
	}
\label{fig:3pcf_subhalo}
\end{figure*} 

We next consider the presence of substructures in a spherical halo.
The standard CDM model predicts that a halo contains abundant
substructures \citep[e.g.][]{1998MNRAS.299..728T, 2004MNRAS.355..819G}.
To examine the effect of substructures in Eqs~(\ref{eq:gggl})
and (\ref{eq:gggl_kappa}), we consider the following mass model:
\beqa
\kappa(\theta, \varphi) = \kappa_{\rm sph}(\theta)
+ \kappa_{\rm sub}(|{\bd \theta} - {\bd \theta}_{\rm sub}|), 
\label{eq:kappa_sub}
\eeqa
where we consider the polar coordinate ${\bd \theta} = (\theta, \varphi)$ 
and ${\bd \theta}_{\rm sub}$ located at the center of subhalo
with respect to the halo center.
In Eq.~(\ref{eq:kappa_sub}), $\kappa_{\rm sph}$ represents
the convergence field for main spherical NFW halo
with mass of $M_{\rm host}$ and concentration $c_{\rm NFW, host}$, while $\kappa_{\rm sub}$ expresses the additional contribution from subhalo with mass of $M_{\rm sub}$, concentration $c_{\rm sub}$
and offset radius of $|{\bd \theta}_{\rm sub}|$.
Here we compute $\kappa_{\rm sub}$ by using the mass density profile proposed in \citet{2003ApJ...584..541H}.
Note that our model of $\kappa_{\rm sub}$ includes 
the truncation of density profile due to tidal stripping 
\MS{\citep[see Section~2.1 in][for detailed expressions]{2015ApJ...799..188S}.}

To compute the HSSC for the model of 
Eq.~(\ref{eq:kappa_sub}), we consider four different subhalo positions,
$\theta_{\rm sub}/\theta_{\rm vir} = 0.2, 0.4, 0.6$ and 0.8 
and fix the other parameters as follows:
$M_{\rm host}=10^{14}\, h^{-1}M_{\odot}$,
$c_{\rm NFW, host}=5.6$,
$M_{\rm sub}=10^{13}\, h^{-1}M_{\odot}$,
$c_{\rm sub}=7$,
and 
the lens redshift of 0.3. 
For a given set of parameters, we generate 300 independent mass models
by randomly setting the azimuthal location of the subhalo,
and then compute the shear fields through Fourier transform.
We calculate the mean three-point correlations as in Eqs~(\ref{eq:gggl})
and (\ref{eq:gggl_kappa}) by averaging over the 300 realizations.
Figure~\ref{fig:3pcf_subhalo} shows the model prediction.
We find that squeezed triangles with $\phi\rightarrow0$ give 
large $\zeta_{h\kappa\kappa}$,
while $\zeta_{h\kappa\kappa}$ is small
when $\theta_{\rm sub}$ is not in the radial bin between
$\theta_{1}$ and $\theta_{2}$.
For instance, we find a significant correlation 
only for $\theta_{1}/\theta_{\rm vir}=\theta_{2}/\theta_{\rm vir}=0.3-0.5$
for $\theta_{\rm sub}/\theta_{\rm vir}=0.4$.
This is expected because 
squeezed triangles with $\theta_{1}\simeq\theta_{2}$ and 
$\phi\rightarrow0$ contain lensing signals 
contributed by the surface mass of substructures.
Since our model of $\kappa_{\rm sub}$ predicts a smaller tidal truncation
when subhalos are closer to the halo centre,
the enhancement in $\zeta_{h\kappa\kappa}$ also becomes smaller at smaller $\theta$ even if the subhalo mass $M_{\rm sub}$ is fixed.
The non-local nature of shear field
reduce $\zeta_{h++}$ for
closed triangles but increase the correlation for more general
triangular configurations.
Note that the presence of subhalos breaks the azimuthal symmetry in
the convergence field, 
making the angular ($\theta, \varphi$) dependence of the three-point correlation
more complicated.

\section{Modelling the off-centering effect on halo-shear-shear correlation}

In this appendix, we examine the effect of galaxy-halo off-centering
on the halo-shear-shear correlation.
To perform stacked lensing analyses, it is necessary to determine
the central position of the foreground lensing halo.
For a galaxy cluster identified in an optical survey, it is often
assumed that the brightest cluster galaxy (BCG) 
is located at the center of its host halo.
Recent observations show, however, that BCGs are not always at the center of their host halos,
but their positions are distributed with angular offsets
from the projected halo centers \citep[e.g.,][]{2010MNRAS.405.2215O, 2012MNRAS.426.2944Z}.
Off-centering of massive galaxies have also been studied
in \citet{2013MNRAS.435.2345H}.

\begin{figure*}
\centering
\includegraphics[width=0.70\columnwidth, bb=0 0 521 512]
{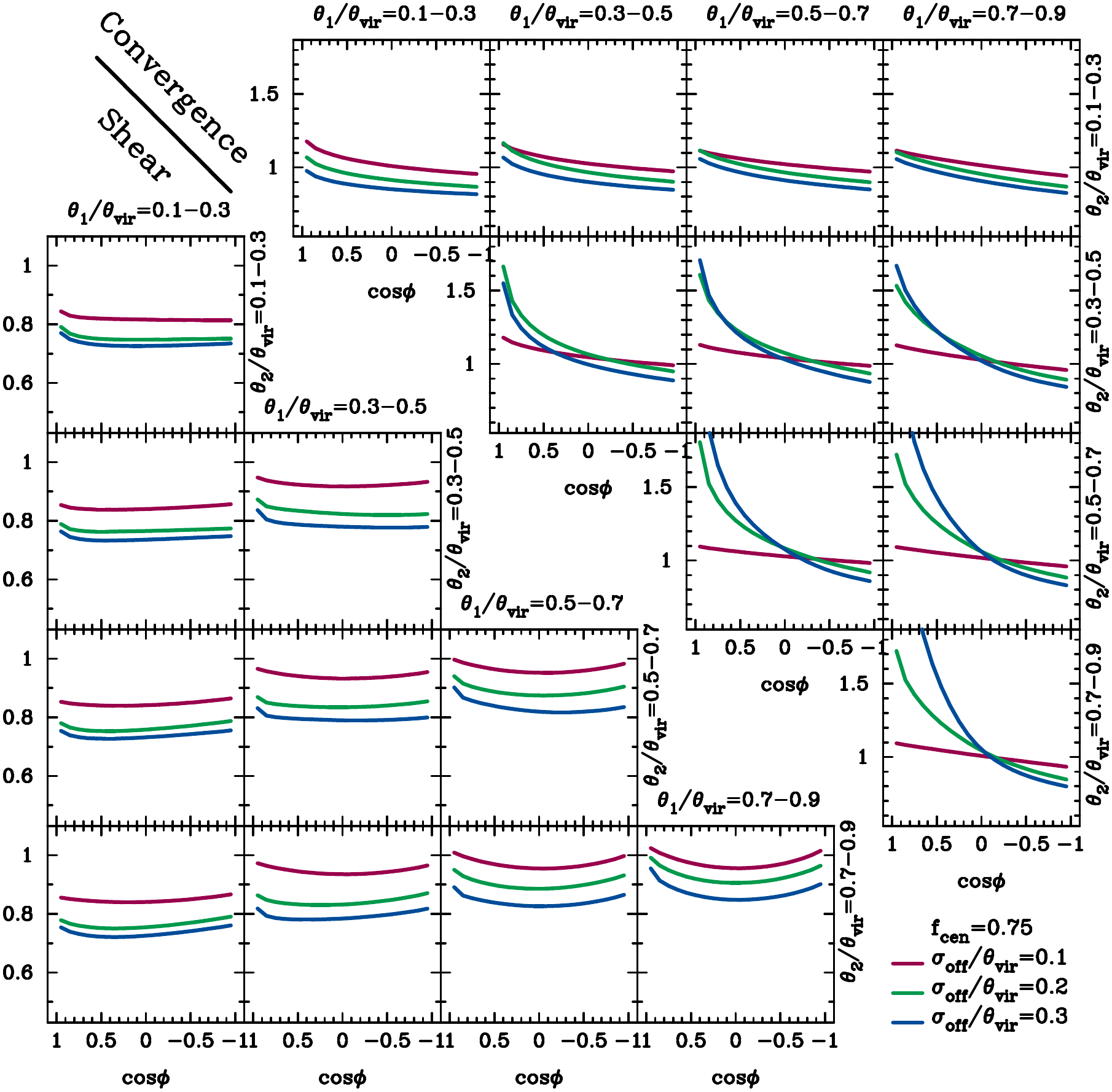}
\caption{
	The off-centering effect on the three-point correlations for spherical halos
	with $M=10^{14}\, h^{-1}M_{\odot}$, $z=0.3$ and $c_{\rm NFW}=5.6$.
	In this figure, we assume that the off-centering probability is given
        as in Eq.~(\ref{eq:Prob_off})
	with $f_{\rm cen}=0.75$. The colored lines show the results with
        three different off-centering
	radius, $\sigma_{\rm off}/\theta_{\rm vir}=0.1, 0.2$ and 0.3,
	where $\theta_{\rm vir}$ is the angular virial radius of the halo.
	The lower triangular panels show the correlation $\zeta_{h++}$,
	while the upper panels are for $\zeta_{h\kappa\kappa}$.
	We normalize the three-point correlations by the respective 
    halo-shear or halo-convergence correlation in the absence of off-centering effects.
	Note that the source redshift is assumed to be 1.
	}
\label{fig:miscenter_sph}
\end{figure*} 

The probability of offset between a reference of halo center and true center 
in stacked lensing analyses is commonly characterized as
\beqa
p(\theta_{\rm off}) = f_{\rm cen} 
+ (1-f_{\rm cen}) \frac{\theta_{\rm off}}{\sigma^2_{\rm off}}
\exp\left(-\frac{\theta^2_{\rm off}}{2\sigma^2_{\rm off}}\right), \label{eq:Prob_off}
\eeqa 
where $\theta_{\rm off}$ represents the offset in angular scale,
$f_{\rm cen}$ is the fraction of objects located at true halo center,
and $\sigma_{\rm off}$ is the scatter in off-centering probability.
To study the impact of possible offsets of the center in stacked analysis,
we first consider the simplest case of spherical NFW halos.
We generate 300 realizations of spherical NFW halos 
with mass $M=10^{14}\, h^{-1}M_{\odot}$ and the concentration
$c_{\rm NFW}=5.6$ at redshift of 0.3 and 
then compute projected density in two-dimensional plane
assuming the central position of these halos follows the probability as in Eq~(\ref{eq:Prob_off}).
Using Eqs.~(\ref{eq:kappa2shear1}) and (\ref{eq:kappa2shear2}),
we obtain the corresponding lensing signals for each off-centered halo in two-dimensional image.
We then perform stacking analyses of three-point correlation 
assuming the stacked position is set to be the center of the image.

Figure~\ref{fig:miscenter_sph} summarizes the results 
from stacked analyses of off-centering spherical NFW halos with the source redshift of 1.
In this figure, we assume $f_{\rm cen}=0.75$ that is broadly consistent with observations,
while varying the scatter of $\sigma_{\rm off}$ as 
$0.1\theta_{\rm vir}$, $0.2\theta_{\rm vir}$, and $0.3\theta_{\rm vir}$.
We find that off-centering effects of a halo-center reference can induce
additional correlation signals even if the shape of halo is exactly spherical.
The effect on the correlation for convergence can show the significant dependence of 
azimuthal angle, but it is completely different from the expectation for the elliptical halos
(see Figure~\ref{fig:3pcf_ellip_halo}).
On the other hand, the off-centering effect can affect 
the amplitude of halo-shear-shear correlation by a factor of $\sim0.7$
and the additional azimuthal dependence is found to be relatively small.

\begin{figure*}
\centering
\includegraphics[width=0.70\columnwidth, bb=0 0 535 512]
{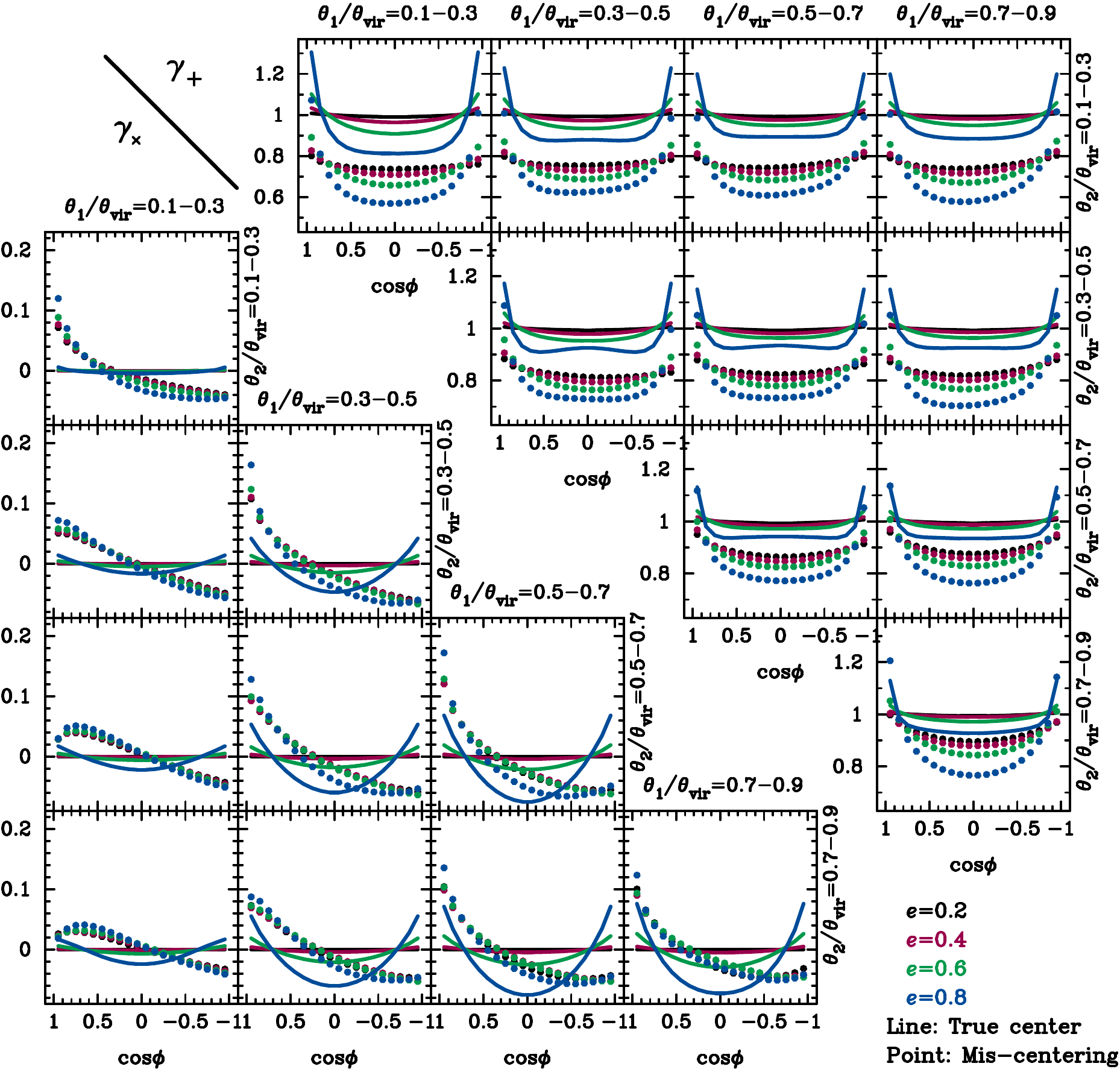}
\caption{
	The off-centering effect of three-point correlation for elliptical halos
	with $M=10^{14}\, h^{-1}M_{\odot}$, $z=0.3$ and $c_{\rm NFW}=5.6$.
	The upper triangular panels show the correlation for tangential shear $\zeta_{h++}$,
	while the lower panels show the correlation for cross shear $\zeta_{h\times\times}$.
	The three-point correlation in each panel is normalized with
    halo-shear correlation without off-centering effects.
	In each panel, the colored points represent the case in the presence of off-centering effects,
	while the lines are for the correlation without off-centering effects.
	The color differences show the different halo ellipticities of $e=0.2$, 0.4, 0.6, and 0.8. 
	Here we assume $f_{\rm cen}=0.75$ and $\sigma_{\rm off}/\theta_{\rm vir}=0.2$
	when including the off-centering effect.
	Note that the source redshift is assumed to be 1.
	}
\label{fig:miscenter_elliptical}
\end{figure*} 

Furthermore, we consider the off-centering of elliptical NFW halos as in
Eq.~(\ref{eq:kappa_ellip}) assuming $f_{\rm cen}=0.75$ 
and $\sigma_{\rm off}/\theta_{\rm vir}=0.2$.
Similarly to the spherical case, we generate 300 realizations of elliptical NFW halos
including the offset of halo positions as in Eq.~(\ref{eq:Prob_off})
and derive the lensing quantities through Fourier transformation with Eqs.~(\ref{eq:kappa2shear1}) and (\ref{eq:kappa2shear2}) in two-dimensional image.
We then compute the corresponding stacked lensing signals by 
setting a reference of halo center to be the center of image.
Figure~\ref{fig:miscenter_elliptical} shows the result of our analyses
of elliptical NFW halos with possible off-centering effects.
The upper triangular panels show the three-point correlation for tangential shear,
while the lower panels for cross shear.
The lines in this figure correspond to the case of $f_{\rm cen}=1$,
while off-centered cases are summarized in the colored points.
The difference of colors represents the different halo ellipticities.
For the tangential shear, we find that the off-centering effect can mainly change the
amplitude of correlation and 
additional azimuthal dependences will be subdominant.
The change of amplitude in $\zeta_{h++}$ can be reasonably approximated as
$A_{0} + A_{1} \cos^2 \phi$, where $A_{0}\simeq0.6-0.8$ and $A_{1}\sim0.1$
for $\theta<\theta_{\rm vir}$ and $e\simlt0.8$.
More importantly, the three-point correlation of cross shear is found to be
less sensitive to halo ellipticities and 
they can be used as a proxy of off-centering effects
(e.g., see the panel for $\theta_{1}/\theta_{\rm vir}=0.1-0.3$ and $\theta_{2}/\theta_{\rm vir}=0.1-0.3$ in Figure~\ref{fig:miscenter_elliptical}).

In a short summary, the off-centering effect of a halo-center reference in stacked analyses
can induce additional correlations, but the major effect on $\zeta_{h++}$ can
be approximated as the change of amplitude alone. Moreover, the degeneracy between
the off-centering effect and underlying correlation induced by halo shapes 
can be broken when the three-point correlation function for cross shear is properly used. 
Assuming $f_{\rm cen}=0.75$ and $\sigma_{\rm off}/\theta_{\rm vir}=0.2$,
the results in Table~\ref{tb:deltachi2} will be modified by 
a factor of $0.7^2-0.8^2 \simeq0.5-0.6$, 
while it seems still possible to detect the halo-shear-shear correlation 
for galaxy clusters and massive galaxies in future imaging surveys.

\end{document}